\documentclass[a4paper,USenglish,cleveref]{lipics-v2021}

\title{Peak Demand Minimization via Sliced Strip Packing}

\author{Max A. Deppert}{Kiel University, Kiel, Germany}{made@informatik.uni-kiel.de}{https://orcid.org/0000-0003-3083-7998}{Research supported by German Research Foundation (DFG) project JA 612/25-1}
\author{Klaus Jansen}{Kiel University, Kiel, Germany}{kj@informatik.uni-kiel.de}{https://orcid.org/0000-0001-8358-6796}{Research supported by German Research Foundation (DFG) project JA 612/25-1}
\author{Arindam Khan}{Department of Computer Science and Automation, Indian Institute of Science, Bengaluru, India}{arindamkhan@iisc.ac.in}{https://orcid.org/0000-0001-7505-1687}{}
\author{Malin Rau}{Universität Hamburg, Hamburg, Germany}{rau@informatik.uni-hamburg.de}{https://orcid.org/0000-0002-5710-560X}{}
\author{Malte Tutas}{Kiel University, Kiel, Germany}{m.tutas@gmx.net}{https://orcid.org/0000-0002-1360-4634}{}

\authorrunning{M.\,Deppert, K.\,Jansen, A.\,Khan, M.\,Rau, M.\,Tutas}

\Copyright{\ } 



\category{} 

\relatedversion{} 

\supplement{}

\nolinenumbers 

\hideLIPIcs  

\EventEditors{John Q. Open and Joan R. Access}
\EventNoEds{2}
\EventLongTitle{42nd Conference on Very Important Topics (CVIT 2016)}
\EventShortTitle{CVIT 2016}
\EventAcronym{CVIT}
\EventYear{2016}
\EventDate{December 24--27, 2016}
\EventLocation{Little Whinging, United Kingdom}
\EventLogo{}
\SeriesVolume{42}
\ArticleNo{23}

\usepackage[dvipsnames]{xcolor}
\usepackage{amsthm}
\usepackage{amsfonts}
\usepackage{mdframed}
\usepackage{graphicx}
\usepackage[colorinlistoftodos]{todonotes}
\usepackage{verbatim} 
\usepackage{enumitem}
\usepackage{fetamont}
\usetikzlibrary{patterns}
\usepackage{caption, setspace, subcaption}
\usepackage{xspace}
\usepackage[nolist]{acronym}
\usepackage{ifthen}
\usepackage{enumitem}
\usepackage{braket}
\usepackage[nolist]{acronym}
\usepackage{tikz}
\usepackage{pgfplots}\pgfplotsset{compat=1.17}
\usepackage{comment}

\usetikzlibrary{shapes.geometric, arrows.meta, graphs, matrix}
\tikzset{%
	>={Latex[width=2mm,length=2mm]},
	base/.style = {rectangle, rounded corners, draw=black,
		minimum width=4cm, minimum height=1cm,
		text centered, font=\sffamily},
	activityStarts/.style = {base, fill=blue!30},
	startstop/.style = {base, fill=red!30},
	activityRuns/.style = {base, fill=subbi!30},
	process/.style = {base, minimum width=2.5cm, fill=orange!15,
		font=\ttfamily},
}
\tikzset{toprule/.style={%
		execute at end cell={%
			\draw [line cap=rect,#1] (\tikzmatrixname-\the\pgfmatrixcurrentrow-\the\pgfmatrixcurrentcolumn.north west) -- (\tikzmatrixname-\the\pgfmatrixcurrentrow-\the\pgfmatrixcurrentcolumn.north east);%
		}
	},
	bottomrule/.style={%
		execute at end cell={%
			\draw [line cap=rect,#1] (\tikzmatrixname-\the\pgfmatrixcurrentrow-\the\pgfmatrixcurrentcolumn.south west) -- (\tikzmatrixname-\the\pgfmatrixcurrentrow-\the\pgfmatrixcurrentcolumn.south east);%
		}
	}
}
\tikzstyle{startstop} = [rectangle, rounded corners, minimum width=3cm, minimum height=1cm,text centered, draw=black, fill=red!30]
\tikzstyle{io} = [trapezium, trapezium left angle=70, trapezium right angle=110, minimum width=3cm, minimum height=1cm, text centered, draw=black, fill=blue!30]
\tikzstyle{process} = [rectangle, minimum width=3cm, minimum height=1cm, text centered, draw=black, fill=orange!30]
\tikzstyle{decision} = [diamond, minimum width=3cm, minimum height=1cm, text centered, draw=black, fill=green!30]
\tikzstyle{arrow} = [thick,->,>=stealth]

\definecolor{subbi}{rgb}{0.53, 0.81, 1}
\definecolor{color2}{rgb}{0.95, 0.53, 1 }
\definecolor{colora3}{rgb}{1, 0.72, 0.53 }
\definecolor{stripColor}{rgb}{0.57, 1, 0.53}
\definecolor{hugeItemColor}{rgb}{1, 0.53, 0.57 }
\definecolor{background}{rgb}{0.6,0.6,0.6}
\newtheorem{thm}{Theorem}
\newtheorem{lem}[thm]{Lemma}
\newtheorem{cor}[thm]{Corollary}

\newcommand{\eps}{\varepsilon}

\newcommand{\opt}{\mathrm{OPT}}
\newcommand{\OPT}{\opt}

\newcommand{\Rsm}{\items_{small}}
\newcommand{\Rla}{\items_{large}}
\newcommand{\Rho}{\items_{hor}}
\newcommand{\Rve}{\items_{ver}}

\newcommand{\height}{e}
\newcommand{\width}{p}

\newcommand{\cTime}{c}
\newcommand{\area}{\mathrm{work}}

\newcommand{\items}{\mathcal{J}}
\newcommand{\res}{\mathrm{res}}

\newcommand{\sched}{\sigma}

\newcommand{\itemsBuck}{\items_{\mathrm{cont}}}
\newcommand{\Cbuck}{C_{\mathrm{cont}}}
\newcommand{\Ctall}{C_{\mathrm{tall}}}
\newcommand{\itemsMove}{\items_{\mathrm{Move}}}
\newcommand{\itemsPar}{\items_{\mathrm{seq}}} 

\newcommand{\Oh}{\mathcal{O}}

\newboolean{BlackAndWhite}
\setboolean{BlackAndWhite}{true}
\definecolor{li}{HTML}{00677C}
\definecolor{wi}{HTML}{8E217D}
\definecolor{ti}{HTML}{F29400}
\definecolor{si}{HTML}{E43117}
\definecolor{vi}{HTML}{39842E}
\definecolor{hi}{HTML}{9B0a7d}

\newcommand{\drawVerticalItem}[5][$ $]{%
	\ifthenelse{\boolean{BlackAndWhite}}{%
		\draw[fill = white!85!black, fill opacity = 0.7] (#2,#3) rectangle node[midway, opacity = 1]{#1} (#4,#5)}{%
		\draw[fill = white!50!vi, fill opacity = 0.7] (#2,#3) rectangle node[midway, opacity = 1]{#1} (#4,#5)}}
\newcommand{\drawVerticalItemRotate}[5][$ $]{%
	\ifthenelse{\boolean{BlackAndWhite}}{%
		\draw[fill = white!85!black, fill opacity = 0.7] (#2,#3) rectangle node[midway, opacity = 1, rotate=90]{#1} (#4,#5)}{%
		\draw[fill = white!50!vi, fill opacity = 0.7] (#2,#3) rectangle node[midway, opacity = 1,rotate=90]{#1} (#4,#5)}}		
\newcommand{\drawTallItem}[5][$ $]{%
	\ifthenelse{\boolean{BlackAndWhite}}{%
		\draw[fill = white!65!black, fill opacity = 0.7] (#2,#3) rectangle node[midway, opacity = 1]{#1} (#4,#5)}{%
		\draw[fill = white!50!ti, fill opacity = 0.7] (#2,#3) rectangle node[midway, opacity = 1]{#1} (#4,#5)}}
\newcommand{\drawLargeItem}[5][$ $]{	\ifthenelse{\boolean{BlackAndWhite}}{%
		\draw[fill = white!55!black, fill opacity = 0.7] (#2,#3) rectangle node[midway, opacity = 1]{#1} (#4,#5)}{%
		\draw[fill = white!50!li, fill opacity = 0.7] (#2,#3) rectangle node[midway, opacity = 1]{#1} (#4,#5)}}
\newcommand{\drawSmallItem}[5][$ $]{	\ifthenelse{\boolean{BlackAndWhite}}{%
		\draw[fill = white!95!black, fill opacity = 0.7] (#2,#3) rectangle node[midway, opacity = 1]{#1} (#4,#5)}{%
		\draw[fill = white!50!si, fill opacity = 0.7] (#2,#3) rectangle node[midway, opacity = 1]{#1} (#4,#5)}}
\newcommand{\drawHorizontalItem}[5][$ $]{	\ifthenelse{\boolean{BlackAndWhite}}{%
		\draw[fill = white!75!black, fill opacity = 0.7] (#2,#3) rectangle node[midway, opacity = 1]{#1} (#4,#5)}{%
		\draw[fill = white!50!hi, fill opacity = 0.7] (#2,#3) rectangle node[midway, opacity = 1]{#1} (#4,#5)}}
\newcommand{\drawJobNoBorder}[5][$ $]{	\ifthenelse{\boolean{BlackAndWhite}}{%
    	\draw[white!90!black, fill, fill opacity = 0.7] (#2,#3) rectangle node[midway, opacity = 1]{#1} (#4,#5)}{%
		\draw[white!50!vi, fill, fill opacity = 0.7] (#2,#3) rectangle node[midway, opacity = 1]{#1} (#4,#5)}}

\newcounter{steplistcount}
\newcounter{steplistcounti}
\newcounter{caselistcount}
\newcounter{caselistcounti}
\renewcommand*\thesteplistcount{\arabic{steplistcount}}
\renewcommand*\thesteplistcounti{\arabic{steplistcounti}}
\renewcommand*\thecaselistcount{\arabic{caselistcount}}
\renewcommand*\thecaselistcounti{\arabic{caselistcounti}}
\newlist{stepList}{description}{2}
\setlist[stepList,1]{%
	before={\setcounter{steplistcount}{0}},
	leftmargin=0em,labelindent=0\parindent,labelsep = \parindent, jobsep=0.1\baselineskip,topsep=0.1\baselineskip,itemsep =\topsep,  listparindent=\parindent,style = sameline
	,font=\normalfont\normalsize\itshape Step~\stepcounter{steplistcount}\thesteplistcount:~ 
}
\setlist[stepList,2]{%
	before={\setcounter{steplistcounti}{0}},%
	leftmargin=0em,labelindent=0\parindent,labelsep = \parindent, jobsep=0.1\baselineskip,topsep=0.1\baselineskip,itemsep =\topsep,  listparindent=\parindent,style = sameline%
	,font=\normalfont\normalsize\itshape Step~ \stepcounter{steplistcounti}\thesteplistcount.\thesteplistcounti:~ 
}

\newlist{caseList}{description}{2}
\setlist[caseList,1]{%
	before={\setcounter{caselistcount}{0}},%
	leftmargin=0em,labelindent=0\parindent,labelsep = \parindent, jobsep=0.1\baselineskip,topsep=0.1\baselineskip,itemsep =\topsep, listparindent=\parindent,style = sameline%
	,font=\normalfont\normalsize\itshape Case~\stepcounter{caselistcount}\thecaselistcount:~
}

\setlist[caseList,2]{%
	before={\setcounter{caselistcounti}{0}},%
	leftmargin=0em,labelindent=0\parindent,labelsep = \parindent, jobsep=0.1\baselineskip,topsep=0.1\baselineskip,itemsep =\topsep,  listparindent=\parindent,style = sameline%
	,font=\normalfont\normalsize\itshape Case~ \stepcounter{caselistcounti}\thecaselistcount.\thecaselistcounti:~ %
}

\newcommand{\jobs}{\mathcal{J}}
\newcommand{\proc}{p}
\newcommand{\energy}{e}
\newcommand{\hmax}{e_{\max}}
\newcommand{\deadline}{D}

\begin{document}
\maketitle

\begin{acronym}
\acro{SliecedStripPacking}[SSP]{Sliced Strip Packing}
\acro{preemptiveProblem}[PPDM]{Preemptive Peak Demand Minimization}
\acro{nonpreemptiveProblem}[NPDM]{Nonpreemptive Peak Demand Minimization}
\acro{mixedProblem}[MPDM]{Mixed Peak Demand Minimization}
\end{acronym}

\begin{abstract}
We study \acf{nonpreemptiveProblem} problem, where we are given a set of jobs, specified by their processing times and energy requirements. The goal is to schedule all jobs within a fixed time period such that the peak load (the maximum total energy requirement at any time) is minimized. 
This problem has recently received significant attention due to its relevance in smart-grids. Theoretically, the problem is related to the  classical strip packing problem (SP).
In SP, a given set of axis-aligned rectangles must be packed into a fixed-width strip, such that the height of the strip is minimized. NPDM can be modeled as strip packing with slicing and stacking constraint: each rectangle may be cut vertically into multiple slices and the slices may be packed into the strip as individual pieces. The stacking constraint forbids solutions where two slices of the same rectangle are intersected by the same vertical line. Nonpreemption enforces the slices to be placed in contiguous horizontal locations (but may be placed at different vertical locations). 

We obtain a $(5/3+\eps)$-approximation algorithm for the problem. 
We also provide an asymptotic efficient polynomial-time approximation scheme (AEPTAS) which generates a schedule for almost all jobs with energy consumption $(1+\eps) \opt$.
The remaining jobs fit into a thin container of height $1$. 
The previous best for NPDM was 2.7 approximation based on FFDH \cite{ranjan2015offline}.
One of our key ideas is providing several new lower bounds on the optimal solution of a geometric packing, which could be useful in other related problems. These lower bounds help us to obtain approximative solutions based on Steinberg's algorithm in many cases.
In addition, we show how to split schedules generated by the AEPTAS into few segments and to rearrange the corresponding jobs to insert the thin  container mentioned above.
\end{abstract}



\section{Introduction}
Recent years have seen a substantial increase in the demand of electricity, due to rapid urbanization, economic growth and new modes of electrical energy consumption (e.g., electric cars).
Traditionally, electricity generation, transmission, and distribution relied on building infrastructure to support the peak load, when the demand for electricity is maximum. 
However, the peak is rarely achieved and thus more demands can be accommodated using the inherent flexibility of scheduling of certain jobs. E.g., energy requirements for HVAC unit, electric vehicle, 
washer and dryer, water heater, etc. can be met with a flexible scheduling of these appliances.
Smart-grids \cite{elecApp2, karbasioun2018asymptotically, siano2014demand} are next-generation cyber-physical systems that couple  digital communication systems on top of the existing grid infrastructure for such efficient utilization of power, e.g., by shifting users’ demand to off-peak hours in order to reduce peak load.

Future smart-grids are expected to obtain demand requirements for a time period and schedule the jobs such that the peak demand is minimized. Recently, this problem has received considerable attention \cite{alamdari2013smart, liu2016optimal,  ranjan2014offline, ranjan2016smart, chakraborty2017efficient}. Each job can also be modeled as a rectangle, with desired power demand as height and required running time as width. This gives a geometric optimization problem where the goal is to pack the  slices of the rectangles into a strip of width as the time period. The goal is to minimize the maximum height of the packing. There is another additional {\em  stacking constraint} requiring that no vertical line may intersect two slices from the same rectangle. 

In this paper, we study this problem known as  \acf{nonpreemptiveProblem}). 
Formally, we are given a set of jobs $\jobs$. 
Each job $j \in \jobs$ has a processing time $\proc(j)\in \mathbb{N}$ (also called width
) and an energy requirement $\energy(j) \in \mathbb{N}$ (also called height
). 
Furthermore, we are given a deadline $\deadline \in  \mathbb{N}$. 
All the jobs are available from the time $0$ and have to be finished before the deadline $\deadline$. 
A schedule $\sched$ of the jobs $\jobs$ assigns each job a starting time $\sched(j) \in \mathbb{N}$ such that it is finished before the deadline, i.e., $\cTime(j) := \sched(j) + \proc(j) \leq \deadline$.
The total energy consumption at a time $\tau \in \{0,\dots,D\}$ is given by $\energy(\tau) := \sum_{j \in \jobs, \sched(j)\leq \tau < \sched(j)+\proc(j)}\energy(j)$.
The objective is to minimize the peak of energy consumption, i.e., minimize $T_{\sched} :=\max_{\tau \in \{0,\dots,D-1\}} \energy(\tau)$.

NPDM can be viewed as a variant of strip packing problem, where we are allowed to slice the rectangles vertically and the slices must be packed in contiguous horizontal positions (but may be placed at different vertical positions). 
In the classical strip packing problem, we are given a set of rectangles as well as a bounded-width strip and the objective is to find a non-overlapping,  axis-aligned packing of all rectangles into the strip so as to minimize the height of the packing. 
A simple reduction from {\sc partition} problem shows a lower bound of 3/2 for polynomial-time approximation for the problem.
In 1980, Baker et al.~\cite{BakerCR80} first gave a 3-approximation algorithm. Later Coffman et al.~\cite{CoffmanGJT80} introduced two simple shelf-based  algorithms: Next Fit Decreasing Height (NFDH), First Fit Decreasing Height (FFDH), with approximation ratios as 3 and 2.7, respectively.
Sleator \cite{Sleator80} gave a 2.5-approximation. 
Thereafter, Steinberg \cite{Steinberg97} and Schiermeyer \cite{Schiermeyer94} independently improved the approximation ratio to 2. Afterwards, Harren and van stee \cite{HarrenS09} obtained a  1.936-approximation. 
The present best approximation is $(5/3+\eps)$, due to Harren et al.~\cite{HarrenJPS12}.

Alamdari et al.~\cite{alamdari2013smart}
studied a variant where we allow preemption of jobs, also known as two-dimensional strip packing with slicing and stacking constraints (2SP-SSC), or preemptive offline cost optimal scheduling problem (P-OCOSP) \cite{ranjan2015offline}. They showed this variant to be  NP-hard and obtained an FPTAS. They also studied several shelf-based algorithms and provide a practical polynomial time algorithm that allows only one preemption per job. 
Ranjan et al.~\cite{ranjan2016smart}  proposed a practical 4/3-approximation  algorithm for this problem. 

For NPDM, Tang et al.~\cite{elecApp2} first proposed a 7-approximation algorithm.  
Yaw et al.~\cite{yaw2014peak} showed that NPDM is NP-hard to approximate within a factor better than $3/2$. They have given a 4-approximation for a special case when all jobs require the same execution time. 
Ranjan et al.~\cite{ranjan2014offline}, have proposed a 3-approximation algorithms for NPDM. 
They \cite{ranjan2015offline} also proposed an FFDH-based $2.7$-approximation algorithm for a mixed variant where some jobs can be preempted and some can not be preempted.

\subsection{Our Contributions}
We obtain improved approximation algorithms for NPDM. Note that the optimal solutions of sliced strip packing/NPDM and strip packing can be quite different. 
In fact, in \cite{blkadek2015contiguous} an example with a ratio 5/4 is presented.  
Thus, the techniques from strip packing does not always translate directly to our problem. 
We exploit the property that, due to slicing, we can separately guess regions ({\em profile}) for packing of jobs with large energy demand  ({\em tall} jobs) and jobs with large time requirements ({\em wide} jobs). We show that we can remove a small amount of jobs with large energy demand so that we can approximately guess the optimal profile of jobs with large processing time so that their starting positions come from a set containing a constant number of values. This helps us to show the existence of a structured solution that we can pack near-optimally using linear programs. This shows the existence of an asymptotic efficient polynomial-time approximation scheme (AEPTAS):
\begin{thm}
For any $\eps > 0$, there is an algorithm that schedules all jobs such that the peak load is bounded by $(1+\eps)\opt+\hmax$. 
The time complexity of this algorithm is bounded by $\Oh(n \log(n)) + 1/\eps^{1/\eps^{\Oh(1/\eps)}}$.
\end{thm}

In fact we show here a slightly stronger results providing a
schedule for almost all jobs $\jobs \setminus C$ with peak load bounded by $(1+\eps)\opt$ plus a schedule for the remaining jobs $C$ with peak load $\hmax$ and schedule length 
$\lambda D$ for a sufficiently  small $\lambda \in [0,1]$.

Using the AEPTAS and Steinberg's algorithm\cite{Steinberg97}, we obtain our main result:
\begin{thm}\label{53approx}
For any $\eps>0$, there is a polynomial-time $(5/3+\eps)$-approximation algorithm for NPDM.
\end{thm}
Previously, in strip packing (and related problems) the lower bound on the optimal height is given based on the height of the tallest job or the total area of all jobs \cite{Steinberg97, Schiermeyer94}. 
One of our main technical contributions  is to show several additional lower bounds on the optimal load. These bounds may be helpful in other related geometric problems. In fact, these can be helpful to simplify some of the analyses of previous algorithms. 
Using these lower bounds, we show, intuitively, if there is a large amount of tall jobs (or wide jobs) we can obtain a good packing using Steinberg's algorithm. Otherwise, we start with the packing from AEPTAS and modify the packing to obtain a packing within $(5/3+\eps)$-factor of the optimal. This repacking utilizes novel insights about the structure of the packing that precedes it, leading to a less granular approach when repacking. 


\subsection{Related Work}
Strip packing has also been studied under asymptotic approximation. The seminal work of Kenyon and Rémila \cite{KenyonR00} provided an APTAS with an additive term $O(\hmax/\eps^2)$, where $\hmax$ is the height of the tallest rectangle. 
The latter additive term was subsequently improved to $\hmax$ by Jansen
and Solis-Oba \cite{JansenS09}.
Pseudo-polynomial time algorithm for strip packing has received recent attention \cite{NadiradzeW16, GalvezGIK16, AdamaszekKPP17, HenningJRS20}. 
Finally, Jansen and Rau \cite{JansenR19}  gave an almost tight pseudo-polynomial time $(5/4+\eps)$-approximation algorithm.
Recently, Galvez et al. \cite{Galvez0AJ0R20} gave a tight $(3/2+\eps)$-approximation algorithm for a special case when all rectangles are skewed (each has either width or height $\le \delta \deadline$, where $\delta \in (0,1]$ is a small constant).

A related problem is non-contiguous multiple organization packing \cite{bougeret2010approximating}, where
 the width of each rectangle represents a demand for a number of concurrent  processors. This is similar to sliced strip packing, however, the slices need to be  horizontally aligned to satisfy concurrency.
Several important scheduling problems are related such as multiple strip packing \cite{JansenR19}, malleable task scheduling \cite{jansen2004scheduling}, parallel task scheduling \cite{DBLP:journals/siamcomp/JansenT10},  moldable task scheduling \cite{jansen20123,DBLP:conf/ipps/JansenL18}, etc.

Several geometric packing problems are well-studied in combinatorial optimization. In two-dimensional bin packing, we are given a set of rectangles and the goal is to pack  all rectangles into the minimum number of unit square bins. 
This well-studied problem \cite{BansalCS09, JansenP14} is known to admit no APTAS \cite{BansalCKS06},  unless P=NP, and the present best approximation ratio is 1.406 \cite{BansalK14}.
Another related problem is two-dimensional geometric knapsack \cite{JansenZ07, GalSocg21}, where each  rectangle has an associated profit and we wish to pack a maximum profit subset of rectangles in a given square knapsack. The present best approximation ratio for the problem is 1.89 \cite{GalvezGHI0W17}.
These problems are also studied under guillotine cuts \cite{BansalLS05, KhanR20, KhanSocg21} where all jobs can be cut out by a recursive sequence of end-to-end cuts.  There are several other important related problems such as maximum independent set of rectangles \cite{AdamaszekHW19}, unsplittable flow on a path \cite{GMW018}, storage allocation problem \cite{MomkeW20}, etc. 
We refer the readers to \cite{CKPT17} for a recent survey of the area.

\section{Preliminaries}

\noindent \textbf{Notation. }
We denote by $\opt(I)$ (or just $\opt$) the optimal energy consumption peak. 
For some set of jobs $\items$ we define $\area(\items) = \sum_{i \in \items}\width(i)\height(i)$, the total processing time as $\width(\items) = \sum_{i\in\items} \width(i)$, as well as the total energy demand as $\height(\items) = \sum_{i\in\items} \height(i)$. 
With the additional notation of $\items_{P(i)} = \set{i \in \items | P(i)}$ as a restriction of $\items$ using the predicate $P$. E.g., we may express the energy demand time of jobs of $\items$ which have a processing time of at least $\deadline/2$ by $\height(\items_{\width(i) \geq \deadline/2})$.
Furthermore, given a set of jobs $\items$, we denote $\width_{\max}(\items) := \max_{i \in \items} \width(i)$ and $\height_{\max}(\items) := \max_{i \in \items} \height(i)$ and write $\height_{\max}$ and $\width_{\max}$ if the set of jobs is clear from the context.
We say a job $i$ that is placed at $\sigma(i)$ overlaps a point in time $\tau$ if and only if $\sigma(i) \leq \tau < \sigma(i) + \width(i)$.
The set $\items(\tau)$ denotes the set of jobs that overlap the point in time $\tau$.
Additionally, we introduce segments $S$ of the schedule which refer to time intervals and container $C$ which can be seen as sub schedules. 
The starting point of a time interval $S$ will be denoted by $\sched(S)$ and its endpoint as $c(S)$.
On the other hand, a container $C$ has a length (time), which is denoted as $\width(C)$, and a bound on the energy demand $\height(C)$.
If these containers are scheduled, the get a start point $\sched(C)$, which is added to the start point of any job scheduled in $C$.

\subparagraph{Steinberg's Algorithm. } 
Steinberg's algorithm \cite{Steinberg97} will be a crucial subroutine in our algorithms. 
\begin{thm}[Steinberg \cite{Steinberg97}]
\label{thm:steinberg}
Steinberg's algorithm packs a set of rectangular objects $\mathcal{R}$ into a rectangular container of height $a$ and width $b$ in polynomial time, if and only if the following inequalities hold:
\begin{equation}
\tag{Steinberg's Cond.}
\hmax \leq a, \ \ \ \ \ \width_{\max} \leq b, \ \ \ \ \  2 \sum_{r \in \mathcal{R}} \height(r)\width(r)\leq a b - (2 \hmax - a)_+(2 \width_{\max} - b)_+,
\end{equation}
where $x_+ = \max\{x,0\}$, $\width_{\max}$ is the maximal width of a rectangle, and $\hmax$ is the maximal height of a rectangle, $\height(r)$ represents the height of a rectangle and $\width(r)$ represents the width of a rectangle.
\end{thm}

\section{Finding a lower bound on \texorpdfstring{$\opt$}{}}
In this section, we first bound the optimal schedule height from below and then use Steinberg's algorithm to handle some cases.
Two obvious lower bounds are the height of the tallest job $\hmax$ and the bound given by the total area of the jobs, i.e., $\opt \geq \max\{\hmax, \area(\items)/\deadline\}$.
Another simple lower bound is the total height of jobs wider than $\deadline/2$, since they cannot be placed next to each other.
This gives us the first lower bound on $\opt$ and we call it $T_1 := \max\{\hmax, \area(\items)/\deadline, \height(\items_{\width(i) >  \deadline/2})\}$.
In the following, we will present three more complex lower bounds.

\subsection{First bound}

\begin{lem}
\label{lem:tallItemsWidth}
It holds that $\width(\items_{\height(i)>(1/3)\opt}) + \width(\items_{\height(i) > (2/3)\opt}) \leq 2\deadline$ and $\width(\items_{\height(i)>(1/2)\opt}) \leq \deadline$.
\end{lem}
\begin{proof}
Note that jobs with height larger than $(1/3)\opt$ cannot intersect the same vertical line as jobs with height larger than $(2/3)\opt$ in an optimal schedule.
Furthermore, each vertical line through an optimal schedule can intersect at most two jobs with height larger than $(1/3)\opt$. 
Moreover, no vertical line can intersect two jobs from the set $\items_{\height(i)>(1/2)\opt}$.
The claim is a consequence.
\end{proof}

\begin{cor}
The smallest value $T$ such that $\width(\items_{\height(i) > (1/3) T}) + \width(\items_{\height(i) > (2/3)T}) \leq 2\deadline$ and $\width(\items_{\height(i) >  T/2}) \leq \deadline$ is a lower bound for $\opt$.
\end{cor}
\begin{proof}
By Lemma \ref{lem:tallItemsWidth}, we know that $\width(\items_{\height(i)>(1/3)\opt}) + \width(\items_{\height(i) > (2/3)\opt}) \leq 2\deadline$
and obviously we have $\width(\items_{\height(i) >  \opt/2}) \leq \deadline$.
Therefore, the smallest value such that $\width(\items_{\height(i) > (1/3) T}) + \width(\items_{\height(i) > (2/3)T}) \leq 2\deadline$ and $\width(\items_{\height(i) >  T/2}) \leq \deadline$ has to be a lower bound on $\opt$.
\end{proof}

Note that we can find this smallest value in $\Oh(n \log n)$ by starting with $T= T_1$ and as long as  $\width(\items_{\height(i)\geq 1/3T}) + \width(\items_{\height(i)\geq 2/3 T}) > 2\deadline$ or $\width(\items_{ \height(i) >  T/2}) > \deadline$ update $T$ as follows:
For $l \in [0,1]$, denote by $\height_{l}$ the height of the smallest job in $\items_{ \height(i) > l\cdot T}$ and set $T := \min\{3 \height_{1/3}, (3/2)  \height_{2/3}, 2 \height_{1/2}\}$.
This iteratively excludes one job from one of the three sets. 
We denote this lower bound as 
$$T_2 := \min \{T | \width(\items_{\height(i)\geq T/3}) + \width(\items_{\height(i)\geq 2T/3}) \leq 2\deadline \wedge \width(\items_{\height(i)\geq T/2}) \leq \deadline\}.$$

\begin{lem}
\label{lem:T2}
Let $w' \in [0,1/2)$. Then
$$\width(\items_{\height(i) > (2/3)T_2}) > (1-w')\deadline \Rightarrow \width(\items_{\height(i) \in ((1/3)T_2, (2/3)T_2]}) \leq 2w' \deadline.$$
\end{lem}
\begin{proof}
We know that $\width(\items_{\height(i)>(1/3)T_2}) + \width(\items_{\height(i) > (2/3)T_2}) \leq 2\deadline$.
Since $\items_{\height(i) > (2/3)T_2} \subseteq \items_{\height(i)>(1/3)T_2}$ and $ \width(\items_{\height(i) > (2/3)T_2})>(1-w')\deadline$ it holds that $\width(\items_{\height(i) \in ((1/3)T_2, (2/3)T_2]}) \leq 2w' \deadline$.
\end{proof}

\subsection{Second bound}
Next, we obtain a bound based on a set of jobs that do not overlap vertically in a given optimal schedule. 
\begin{lem}
\label{lem:hightOfWideItems1}
Consider an optimal schedule and let $\itemsPar$ be a set of jobs such that no pair of jobs $i,i'\in \itemsPar$ overlaps vertically, i.e., $\sched(i) +\width(i) \leq  \sched(i')$ or $ \sched(i') +\width(i') \leq  \sched(i)$.
Furthermore, define $\items_w := \items_{\width(i) > (\deadline -\width(\itemsPar)/2)}\setminus \itemsPar$.
Then there exists a vertical line through the schedule that intersects a job in $\itemsPar$ and all the jobs in $ \items_w$.
\end{lem}
\begin{proof}
First note that $(\deadline -\width(\itemsPar)/2) \geq \deadline/2$.
Consider the vertical strip between $\width(\itemsPar)/2$ and $(\deadline -\width(\itemsPar))/2$. 
Each job in $\items_w$ completely overlaps this strip.
Furthermore, either the strip itself contains a job in $\itemsPar$, in which case the claim is trivially true, 
or on each position on both sides of the strip there is a job from $\itemsPar$.
Assume the later case.
Since the jobs in $\items_w$ have a width strictly larger than $(\deadline -\width(\itemsPar)/2)$, there exists an $\sigma >0$ such that the vertical line at $(\deadline -\width(\itemsPar)/2) + \sigma$ as well is overlapped by all the jobs in this set.
Since this line intersects also a job from the set $\itemsPar$, the claim follows.
\end{proof}

\begin{cor}
\label{lem:LowerBoundWideTall}
Let $\itemsPar$ be a set of jobs such that $\width(\itemsPar) \leq \deadline$ and consider $\items_w := \items_{\width(i) > \deadline-\width(\itemsPar)/2}\setminus \itemsPar$. Furthermore let $i_{\bot} \in \itemsPar$ be the job with the smallest height.
Then it holds that 
$\min\{\height(i_{\bot}) + \height(\items_w),2\height(i_{\bot})\} \leq \opt$.
\end{cor}
\begin{proof}
Consider an optimal solution.
We have to consider two cases.
In the first case, two jobs from the set $\itemsPar$ intersect the same vertical line. 
In this case, $2\height(i_{\bot})$ is obviously a lower bound on $\OPT$.

On the other hand, if in any optimal schedule there does not exist a pair of jobs from $\itemsPar$ that overlap the same vertical line, we know by Lemma \ref{lem:hightOfWideItems1} that there exists a job in $\itemsPar$ that overlaps with all the jobs in $\items_w$ and therefore $\opt \geq \height(i_{\bot}) + \height(\items_w)$ in this case.
\end{proof}

From Corollary \ref{lem:LowerBoundWideTall}, we derive a lower bound on $\opt$. 
For a given $k \in [n]$, we define $\items_k$ to be the set of the $k$ jobs with the largest energy demand in $\items$ and  $\items_k'$ to be the set of the $k$ jobs with the largest energy demand in $\items \setminus \items_{\width(i) > \deadline/2}$. Let  $i_k$ and $i_k'$ be the jobs with the smallest height in $\items_k$ and $\items_k'$, respectively.
We define: 
$$T_{3,a} := \max \{\min\{\height(i_k) + \height(\items_{\width(i) > \deadline-\width(\items_k)/2} \setminus \items_k),2\height(i_k)\}|k \in \{1,\dots,n\}, \width(\items_k)\leq \deadline\},$$
$$\text{and } T_{3,b} := \max \{\min\{\height(i_k') + \height(\items_{\width(i) > \deadline-\width(\items_k')/2}),2\height(i_k')\}|k \in \{1,\dots,n\}, \width(\items_k')\leq \deadline\},$$
and finally $T_3= \max\{T_{3,a},T_{3,b}\}$.
Note that $\items_k'$ and $\items_{\width(i) > \deadline-\width(\items_k')/2}$ are disjoint, since $\items_k'$ contains only jobs with width at most $\deadline/2$ and $\items_{\width(i) > \deadline-\width(\items_k')/2}$ contains only jobs width width larger than $\deadline/2$, and hence, by Corollary \ref{lem:LowerBoundWideTall}, $T_3$ is a lower bound on $\opt$.

For this lower bound, we prove the following property.

\begin{lem}
\label{lem:totalHeightOfWideItems}
Let $T = \max\{T_1,T_2,T_3\}$, $w \in (0,1/2)$ and $h \in (1/2,1]$ as well as $\items_h := \items_{\height(i) \geq h T}$ and 
$\items_w := \items_{\width(i) > (1/2+w/2)\deadline}\setminus \items_h$. %
It holds that
    $$\width(\items_h) \geq (1-w)\deadline \Rightarrow \height(\items_w) \leq (1-h)T.$$
\end{lem}
\begin{proof}
Since $T\geq T_2$, it holds that $\width(\items_h) \leq \deadline$. By construction of $T_3$ for each job $j \in \items_h$, it holds that $$\height(j) + \height(\items_{\width(i) > \deadline -\width(\items_{\height(i) \geq \height(j) })/2} \setminus \items_{\height(i) \geq \height(j) }) \leq T_3,$$ 
because $2\height(j) > T_3$ (and $T_3 \geq \min\{2\height(j), \height(j) + \height(\items_{\width(i) > \deadline-\width(\items_{\height(i) \geq \height(j) })/2} \setminus \items_{\height(i) \geq \height(j) })\}$).
Furthermore, note that $\items_h = \items_{\height(i) \geq \height(j) }$ for the job $j$ with the smallest height in $\items_h$.

Therefore, if $\width(\items_h) \geq (1-w)\deadline$, it holds that $\items_{\width(i) > \deadline-(1-w)\deadline/2} \subseteq \items_{\width(i) > \deadline-\width(\items_h)/2}$ and hence, 
\begin{align*}
    h T  + \height(\items_w) = h T  + \height(\items_{\width(i) > \deadline-(1-w)\deadline/2} \setminus \items_h) \leq \height(j) + \height(\items_{\width(i) > \deadline-\width(\items_h)/2} \setminus \items_h) \leq T_3 \leq T.
\end{align*}
\end{proof}

\begin{lem}
\label{lem:totalHeightOfWideItems2}
Let $T = \max\{T_1,T_2,T_3\}$, $w \in (1/2,1]$ and $h \in (1/2,1]$ as well as $\items_w := \items_{\width(i) \geq w \deadline}$ and 
$\items_h := \items_{\height(i) > h T}\setminus \items_{\width(i) > \deadline/2}$. %
It holds that
$$ \height(\items_w) > (1-h) T \Rightarrow \width(\items_h) \leq 2(1-w )\deadline.$$
\end{lem}
\begin{proof}
Let $\height(\items_w) > (1-h) T$. 
Since for each job in $\items_h $ it holds that $\height(i) > T/2 \geq T_{3,b}/2$,
by definition of $T_{3,b}$, for each $j \in \items_h$ it holds that 
$$ \height(j) + \height(\items_{\width(i) > \deadline-\width(\items_{\height(i) \geq \height(j)})/2}) \leq T.$$ 
Therefore for the smallest job $j \in \items_h$, it holds that
$$ \height(j) + \height(\items_{\width(i) > \deadline - \width(\items_h)/2 })  \leq T.$$

For contradiction assume that $\width(\items_h) > 2(1-w)\deadline$. Note that in this case $\deadline - \width(\items_h)/2 < w \deadline$ and hence 
$$ \height(\items_{\width(i) > \deadline - \width(\items_h)/2 }) \geq \height(\items_{\deadline}) > (1-h) T.$$
As a consequence 
$$ \height(j) + \height(\items_{\width(i) > \deadline - \width(\items_h)/2 }) > hT + (1-h) T = T,$$
a contradiction.
\end{proof}

\subsection{Third bound}
 
\begin{lem}
\label{lem:hightOfWideItems2}
Consider an optimal schedule and let $\itemsPar \subseteq \items$ be a set of jobs such that no pair of jobs $j,j'\in \itemsPar$ overlaps vertically, i.e., $ \sched(j) +\width(j) \leq  \sched(j')$ or $ \sched(j') +\width(j') \leq  \sched(j)$.
Let $\items_{\deadline} \subseteq \items_{\width(j) > (\max\{\deadline -\width(\itemsPar),\deadline/2\})} \setminus \itemsPar$.
Then there exists a vertical line through the schedule that intersects a job in $\itemsPar$ and a subset $ \items_{W'} \subseteq \items_{\deadline}$ with $\height(\items_{\deadline}') \geq \height(\items_{\deadline})/2$.
\end{lem}
\begin{proof}
First, we consider the trivial cases.
If a job from $\itemsPar$ overlaps the vertical line at $\deadline/2$ the claim is trivially true, since all the jobs from $\items_{\deadline}$ overlap $\deadline/2$.
On the other hand, if all the jobs in $\itemsPar$ are left or right of $\deadline/2$, it holds that $\width(\itemsPar) \leq \deadline/2$ and one of the jobs has a distance of at most $\deadline/2 - \width(\itemsPar)$ from $\deadline/2$.
This job has to be overlapped by all the jobs from $\items_{\deadline}$ since they have a width larger than $\deadline -\width(\itemsPar)$.

Otherwise, consider the vertical line $L_l$ through the right border of the rightmost job from $\itemsPar$ that is left of $\deadline/2$ and the vertical line $L_r$ through the left border of the leftmost job from $\itemsPar$ that is right of $\deadline/2$.
Note that $L_l$ and $L_r$ have a distance of at most $(\deadline-\width(\itemsPar))$.
Consider the set $\items_{\deadline,l} \subseteq \items_{\deadline}$ that is intersected by the vertical line $L_l$.
Note that the residual jobs in $\items_{\deadline,r} :=\items_{\deadline} \setminus \items_{\deadline,l}$ all overlap the vertical line at $L_l + (\deadline -\width(\itemsPar)) \geq L_r$ and hence $L_r$ as well.
Since $\items_{\deadline,r} \cup \items_{\deadline,l} =  \items_{\deadline}$, one of the two sets has a height of at least $\height(\items_{\deadline})/2$.
Finally, note that there exists a small enough $\sigma >0$ such that $L_l -\sigma$ and $L_r+ \sigma$ overlap the same set of wide jobs as $L_l$ and $L_r$ as well as the corresponding job in $\itemsPar$.
\end{proof}

\begin{cor}
\label{lem:LowerBoundT4}
Let $\itemsPar$ be a set of jobs such that $\width(\itemsPar) \leq \deadline$ and consider $\items_{\deadline} := \items_{\width(i) > (\max\{\deadline -\width(\itemsPar),\deadline/2\})} \setminus \itemsPar$. Furthermore let $i_{\bot} \in \itemsPar$ be the job with the smallest height.
Then it holds that 
$\min\{\height(i_{\bot}) + \height(\items_\deadline)/2,2\height(i_{\bot})\} \leq \opt$.
\end{cor}
\begin{proof}
Consider an optimal solution.
We have to consider two cases.
In the first case, two jobs from the set $\itemsPar$ intersect the same vertical line. 
In this case, $2\height(i_{\bot})$ is obviously a lower bound on $\OPT$.

On the other hand, if in any optimal schedule there does not exist a pair of jobs from $\itemsPar$ that overlap the same vertical line, we know by Lemma \ref{lem:hightOfWideItems2} that there exists a job in $\itemsPar$ that overlaps with a set $\items_{\deadline}' \subseteq \items_{\deadline}$ such that $\height(\items_{\deadline}') \geq \height(\items_{\deadline})/2$ and therefore $\opt \geq \height(i_{\bot}) + \height(\items_{\deadline})/2$ in this case.
\end{proof}

Define $\items_k$ as the set of the $k$ jobs with largest energy demand,  $\items_{\deadline,k} := \items_{\width(i) > (\max\{\deadline -\width(\items_k),\deadline/2\})} \setminus \items_k$. Let $i_k$ be the smallest height job in $\items_k$, then define
\[
T_4 := \max\{\min\{2\height(i_k),\height(i_k) + \height(\items_{\deadline,k})/2\}| k \in \{1,\dots,n\}, \width(\items_k)\leq \deadline \}.
\]
By Corollary \ref{lem:LowerBoundT4}, $T_4$ is a lower bound for $\opt$.

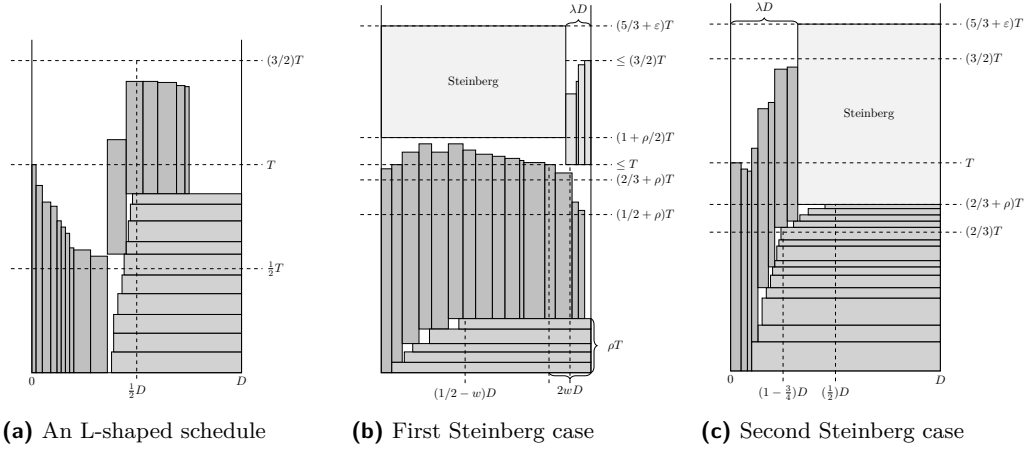
\begin{figure}
    \centering
    \begin{subfigure}[t]{.32\textwidth}
        \centering
        \resizebox{0.98\textwidth}{!}{
        \begin{tikzpicture}

            \pgfmathsetmacro{\w}{6}
            \pgfmathsetmacro{\h}{6}
            \pgfmathsetmacro{\hh}{2/3*\h - 0.26*\h}
            \draw[white] (-0.1*\w,-0.15*\h) rectangle (1.45*\w,5*\h/3 +0.1*\h);

            \drawTallItem{0.00*\w}{0.5*\h-0.5*\h}{0.02*\w}{1.5*\h-0.5*\h};
            \drawTallItem{0.02*\w}{0.6*\h-0.6*\h}{1.0*\w-0.95*\w}{1.5*\h-0.6*\h};
            \drawTallItem{0.05*\w}{0.68*\h-0.68*\h}{0.06*\w+0.03*\w}{1.5*\h-0.68*\h};
            \drawTallItem{0.09*\w}{0.7*\h-0.7*\h}{0.97*\w+0.09*\w-0.94*\w}{1.5*\h-0.7*\h};
            \drawTallItem{0.12*\w}{0.77*\h-0.77*\h}{0.08*\w+0.06*\w}{1.5*\h-0.77*\h};
            \drawTallItem{0.14*\w}{0.8*\h-0.8*\h}{0.94*\w+0.14*\w - 0.92*\w}{1.5*\h-0.8*\h};
            \drawTallItem{0.16*\w}{0.83*\h-0.83*\h}{0.10*\w+0.08*\w}{1.5*\h-0.83*\h};
            \drawTallItem{0.18*\w}{0.9*\h-0.9*\h}{0.92*\w+0.18*\w - 0.90*\w}{1.5*\h-0.9*\h};
            \drawTallItem{0.20*\w}{0.91*\h-0.91*\h}{0.18*\w+0.1*\w}{1.5*\h-0.91*\h};
            \drawTallItem{0.28*\w}{0.94*\h-0.94*\h}{0.26*\w+0.1*\w}{1.5*\h-0.94*\h};
            \drawTallItem{0.36*\w}{0.95*\h-0.38*\h}{0.90*\w+0.36*\w - 0.81*\w}{1.5*\h-0.38*\h};
            \drawTallItem{0.45*\w}{0.96*\h-0.1*\h}{0.34*\w+0.19*\w}{1.5*\h-0.1*\h};
            \drawTallItem{0.53*\w}{0.96*\h-0.1*\h}{0.81*\w+0.53*\w-0.74*\w}{1.5*\h-0.1*\h};
            \drawTallItem{0.60*\w}{0.965*\h-0.105*\h}{0.74*\w+0.60*\w - 0.65*\w}{1.5*\h-0.105*\h};
            \drawTallItem{0.69*\w}{0.98*\h-0.12*\h}{0.38*\w+0.69*\w - 0.34*\w}{1.5*\h-0.12*\h};
            \drawTallItem{0.73*\w}{0.985*\h-0.125*\h}{0.65*\w+0.73*\w - 0.63*\w}{1.5*\h-0.125*\h};
            
            \drawHorizontalItem{0.38*\w}{0.0*\h}{\w}{0.1*\h};
            \drawHorizontalItem{0.39*\w}{0.19*\h}{\w}{0.1*\h};
            \drawHorizontalItem{0.39*\w}{0.28*\h}{\w}{0.19*\h};
            \drawHorizontalItem{0.41*\w}{0.28*\h}{\w}{0.38*\h};
            \drawHorizontalItem{0.43*\w}{0.47*\h}{\w}{0.38*\h};
            \drawHorizontalItem{0.44*\w}{0.47*\h}{\w}{0.57*\h};
            \drawHorizontalItem{0.45*\w}{0.63*\h}{\w}{0.57*\h};
            \drawHorizontalItem{0.46*\w}{0.63*\h}{\w}{0.73*\h};
            \drawHorizontalItem{0.47*\w}{0.81*\h}{\w}{0.73*\h};
            \drawHorizontalItem{0.48*\w}{0.81*\h}{\w}{0.86*\h};
            
            \draw (0,3*\h/2+0.1*\h) -- (0,0) node[below] {$0$} -- (\w,0) node[below] {$\deadline$} -- (\w,3*\h/2+0.1*\h);
            \draw [dashed] (-0.1*\w,3*\h/2) -- (1.1*\w,3*\h/2) node[right] {$(3/2)T$};
            \draw [dashed] (-0.1*\w,\h) -- (1.1*\w,\h) node[right] {$T$};
            \draw [dashed] (-0.1*\w,.5*\h) -- (1.1*\w,.5*\h) node[right] {$\frac{1}{2}T$};
            
            \draw [dashed] (0.5*\w,3*\h/2) -- (0.5*\w,-0.2) node[below] {$\frac{1}{2}\deadline$};
        \end{tikzpicture}
        }
        \caption{An L-shaped schedule}
        \label{fig:AnL-shapedPacking}
    \end{subfigure}
    \begin{subfigure}[t]{.32\textwidth}
        \centering
        \resizebox{0.98\textwidth}{!}{
        \begin{tikzpicture}
            
        \pgfmathsetmacro{\w}{6}
        \pgfmathsetmacro{\h}{6}
        \pgfmathsetmacro{\hh}{2/3*\h - 0.26*\h}
        \pgfmathsetmacro{\r}{ 0.26}
        \pgfmathsetmacro{\l}{ 0.88}
        \draw[white] (-0.1*\w,-0.15*\h- 0.26*\h) rectangle (1.45*\w,5*\h/3 +0.1*\h- 0.26*\h);
        
        \drawTallItem{0}{-0.26*\h}{0.05*\w}{0.98*\h-0.26*\h};
        \drawTallItem{0.05*\w}{-0.21*\h}{0.1*\w}{0.95*\h-0.21*\h};
        \drawTallItem{0.1*\w}{ -0.12*\h}{0.18*\w}{0.92*\h -0.12*\h};
        \drawTallItem{0.18*\w}{ -0.05*\h}{0.24*\w}{0.89*\h -0.05*\h};
        \drawTallItem{0.24*\w}{ -0.05*\h}{0.32*\w}{0.85*\h -0.05*\h};
        \drawTallItem{0.32*\w}{0}{0.39*\w}{0.84*\h};
        \drawTallItem{0.39*\w}{0}{0.45*\w}{0.81*\h};
        \drawTallItem{0.45*\w}{0}{0.53*\w}{0.79*\h};
        \drawTallItem{0.53*\w}{0}{0.59*\w}{0.78*\h};
        \drawTallItem{0.59*\w}{0}{0.66*\w}{0.77*\h};
        \drawTallItem{0.66*\w}{0}{0.68*\w}{0.76*\h};
        \drawTallItem{0.68*\w}{0}{0.78*\w}{0.75*\h};
        \drawTallItem{0.78*\w}{0}{0.83*\w}{0.74*\h};
        \drawTallItem{0.83*\w}{0}{0.91*\w}{0.70*\h};
        \drawTallItem{0.91*\w}{0}{0.94*\w}{0.56*\h};
        \drawTallItem{0.94*\w}{0}{0.97*\w}{0.52*\h};
        
        \begin{scope}[yshift = -1.26*\h cm]
        \drawHorizontalItem{0.05*\w}{1.0*\h}{\w}{1.05*\h};
        \drawHorizontalItem{0.11*\w}{1.05*\h}{\w}{1.1*\h};
        \drawHorizontalItem{0.15*\w}{1.1*\h}{\w}{1.14*\h};
        \drawHorizontalItem{0.23*\w}{1.14*\h}{\w}{1.21*\h};
        \drawHorizontalItem{0.37*\w}{1.21*\h}{\w}{1.26*\h};
        \draw [decorate,decoration={brace,amplitude=5pt,mirror,raise=0pt},yshift=0pt]
        (\w,\h) -- (\w,1.26*\h) node [black,midway,xshift=0.7cm] {
        $\rho T$};
        \end{scope}

        \draw[fill = white!95!black] (0*\w,\h-0.26*\h/2) rectangle node[midway]{Steinberg} (0.88*\w,5*\h/3 - 0.26*\h);

        \drawVerticalItem{0.97*\w}{\h -\r*\h}{\w}{\h -\r*\h+0.5*\h};
        \drawVerticalItem{0.94*\w}{\h -\r*\h}{0.97*\w}{\h -\r*\h+0.48*\h};
        \drawVerticalItem{0.93*\w}{\h -\r*\h}{0.94*\w}{\h -\r*\h+0.40*\h};
        \drawVerticalItem{\l*\w}{\h -\r*\h}{0.93*\w}{\h -\r*\h+0.34*\h};

        \draw (0,5*\h/3 +0.1*\h-0.26*\h) -- (0,-0.26*\h) -- (\w,-0.26*\h)-- (\w,5*\h/3 +0.1*\h-0.26*\h);
        \draw[dashed] (-0.1*\w,\h/2) --(1.1*\w,\h/2) node[right]{$(1/2+\rho)T$};
        \draw[dashed] (\l*\w,3*\h/2-\r*\h) --(1.1*\w,3*\h/2-\r*\h) node[right]{$\leq (3/2)T$};
        \draw[dashed] (-0.1*\w,2*\h/3) --(1.1*\w,2*\h/3) node[right]{$ (2/3+\rho)T$};
        \draw[dashed] (-0.1*\w,\h-0.26*\h) --(1.1*\w,\h-0.26*\h) node[right]{$\leq T$};
        \draw[dashed] (-0.1*\w,\h-0.26*\h/2) --(1.1*\w,\h-0.26*\h/2) node[right]{$(1+\rho/2)T$};
        \draw[dashed] (-0.1*\w,5*\h/3 - 0.26*\h) --(1.1*\w,5*\h/3 - 0.26*\h) node[right]{$(5/3+\varepsilon)T$};
        
        
        \draw[dashed] (0.9*\w,-0.05*\h-0.26*\h) --(0.9*\w,\h-0.26*\h);
        \draw[dashed] (0.8*\w,-0.05*\h-0.26*\h)  --(0.8*\w,\h-0.26*\h);
        \draw[dashed] (0.4*\w,-0.05*\h - 0.26*\h)node[below]{$(1/2-w)\deadline$}  --(0.4*\w,0);

        \draw [decorate,decoration={brace,amplitude=5pt,mirror,raise=0pt},yshift=0pt]
        (0.8*\w,0*\h- 0.26*\h) -- (\w,0*\h- 0.26*\h) node [black,midway,yshift=-0.5cm] {
        $2w \deadline$};
        
        \draw [decorate,decoration={brace,amplitude=5pt,mirror,raise=0pt},yshift=0pt]
        (\w,5*\h/3-0.26*\h) -- (0.88*\w,5*\h/3-0.26*\h) node [black,midway,yshift=0.5cm] {
        $\lambda \deadline$};
        \end{tikzpicture}
        }
        \caption{First Steinberg case}
        \label{fig:Steinberg1}
    \end{subfigure}
    \begin{subfigure}[t]{.32\textwidth}
        \centering
        \resizebox{.98\textwidth}{!}{
        \begin{tikzpicture}
        \pgfmathsetmacro{\w}{6}
        \pgfmathsetmacro{\h}{6}
        \pgfmathsetmacro{\hh}{2/3*\h - 0.26*\h}
        \draw[white] (-0.1*\w,-0.15*\h) rectangle (1.45*\w,5*\h/3 +0.1*\h);
        \draw (0,5*\h/3 +0.1*\h) -- (0,0) node[below] {$0$} -- (\w,0) node[below] {$\deadline$} -- (\w,5*\h/3 +0.1*\h);

        \draw [decorate,decoration={brace,amplitude=5pt,raise=0pt},yshift=0pt]
        (0,5*\h/3) -- (0.32*\w,5*\h/3) node [black,midway,yshift=0.5cm] {
        $\lambda \deadline$};
        
        \drawHorizontalItem{\w}{  0.00*\h}{0.1*\w}{0.14*\h};
        \drawHorizontalItem{\w}{  0.14*\h}{0.13*\w}{0.22*\h};
        \drawHorizontalItem{\w}{  0.22*\h}{0.15*\w}{0.35*\h};
        \drawHorizontalItem{\w}{  0.35*\h}{0.17*\w}{0.40*\h};
        \drawHorizontalItem{\w}{  0.4*\h}{0.19*\w}{0.46*\h};
        \drawHorizontalItem{\w}{  0.46*\h}{0.2*\w}{0.50*\h};
        \drawHorizontalItem{\w}{  0.5*\h}{0.21*\w}{0.53*\h};
        \drawHorizontalItem{\w}{  0.53*\h}{0.22*\w}{0.6*\h};
        \drawHorizontalItem{\w}{  0.6*\h}{0.23*\w}{0.63*\h};
        \drawHorizontalItem{\w}{  0.63*\h}{0.24*\w}{0.69*\h};
        \drawHorizontalItem{\w}{  0.69*\h}{0.3*\w}{0.72*\h};
        \drawHorizontalItem{\w}{  0.72*\h}{0.33*\w}{0.75*\h};
        \drawHorizontalItem{\w}{  0.75*\h}{0.37*\w}{0.78*\h};
        \drawHorizontalItem{\w}{  0.78*\h}{0.45*\w}{0.8*\h};
        
        \drawTallItem{0}{0*\h}{0.05*\w}{1*\h};
        \drawTallItem{0.05*\w}{0*\h}{0.08*\w}{0.97*\h};
        \drawTallItem{0.08*\w}{0.0*\h}{0.1*\w}{0.96*\h};
        \drawTallItem{0.1*\w}{0.0*\h+0.14*\h}{0.13*\w}{0.93*\h+0.14*\h};
        \drawTallItem{0.13*\w}{0.4*\h}{0.18*\w}{0.86*\h+0.4*\h};
        \drawTallItem{0.18*\w}{0.5*\h}{0.21*\w}{0.79*\h+0.5*\h};
        \drawTallItem{0.21*\w}{0.69*\h}{0.27*\w}{0.76*\h+0.69*\h};
        \drawTallItem{0.27*\w}{0.72*\h}{0.32*\w}{0.74*\h+0.72*\h};
        \draw[fill = white!95!black] (0.32*\w,0.8*\h) rectangle node[midway]{Steinberg} (\w,5*\h/3 );
        \draw[dashed] (0.25*\w,-0.05*\h)node[below]{$(1-\frac{3}{4})\deadline$}  --(0.25*\w,0.69*\h);
        \draw[dashed] (0.5*\w,-0.05*\h)node[below]{$(\frac{1}{2})\deadline$}  --(0.5*\w,0.8*\h);
        \draw[dashed] (-0.1*\w,4*\h/5) --(1.1*\w,4*\h/5) node[right]{$(2/3+\rho)T$};
        \draw[dashed] (-0.1*\w,2*\h/3) --(1.1*\w,2*\h/3) node[right]{$(2/3)T$};
        \draw[dashed] (-0.1*\w,3*\h/2) --(1.1*\w,3*\h/2) node[right]{$(3/2)T$};
        \draw[dashed] (-0.1*\w,\h) --(1.1*\w,\h) node[right]{$T$};
        \draw[dashed] (-0.1*\w,5*\h/3) --(1.1*\w,5*\h/3) node[right]{$(5/3+\varepsilon)T$};
        \end{tikzpicture}
        }
        \caption{Second Steinberg case}
        \label{fig:Steinberg2}
    \end{subfigure}
    \caption{Subfigure \textbf{\ref{fig:AnL-shapedPacking}} shows one possible L-shaped schedule, where $\itemsPar$ contains all the jobs with height larger than $T/2$ and $\items_{\deadline}$ contains all the jobs with width larger than $\deadline/2$.
    Subfigure \textbf{\ref{fig:Steinberg1}} shows a schedule in the case that $\width(\items_{\height(i) > (2/3)T}) \geq (1-w)\deadline$. Subfigure \textbf{\ref{fig:Steinberg2}} shows a schedule in the case that $\height(\items_{\width(i) > (3/4)\deadline}) \geq (2/3)T$.
    }
\end{figure}

Given two disjoint sets of jobs $\items_{seq}$ and $\items_{\deadline}$, we say they are placed \emph{L-shaped}, if the jobs $i \in \items_{\deadline}$ are 
placed such that $ \sched(i)+\width(i) = \deadline$, while the jobs in $\items_{seq}$ are sorted by height and placed left-aligned tallest to the left, see Figure \ref{fig:AnL-shapedPacking}.

\begin{lem}
\label{lem:LPackingT4}
Let $T = \max\{T_1,T_2,T_3,T_4\}$. If we place $\itemsPar := \items_{\height(i) > T/2}$ and $\items_{\deadline} := \items_{\width(i) > \deadline/2} \setminus \itemsPar$ L-shaped, the schedule has a height of at most $T+\height(\items_{\deadline} )/2 \leq (3/2)\opt$.
\end{lem}
\begin{proof}
Consider a vertical line $L$ through the generated schedule. 
If $L$ does not intersect a job from $\itemsPar$, the intersected jobs have a height of at most $\height(\items_{\deadline}) \leq T$.
Otherwise, let $i_L \in \itemsPar$ and $\items_{W,L} \subseteq \items_{\deadline}$ be the jobs intersected by $L$ and define $\items_{\mathrm{seq},L} := \items_{\height(i) \geq \height(i_L)}$.
Note that by definition of the schedule, it holds that $\items_{W,L} \subseteq \items_{\width(i) > (\max\{\deadline -\width(\items_{\mathrm{seq},L}),\deadline/2\})} \setminus \items_{\mathrm{seq},L}$.
Since $\height(i_L) > T_4/2$, it holds that 
$T_4 \geq \height(i_L)  + \height(\items_{W,L})/2,$
by definition of $T_4$.
As a consequence $\height(i_L)  + \height(\items_{W,L}) \leq T + \height(\items_{W,L})/2 \leq T+\height(\items_{\deadline} )/2 \leq(3/2) \opt$.
\end{proof}

\subsection{First Steinberg Case}

\begin{thm}
[First Steinberg Case]
\label{thm:firstStein}
Let $T := \max\{T_1,T_2,T_3,T_4\}$ be the lower bound on $\opt$ as defined above and $w \leq (3/4)\eps$.

If $\width(\items_{\height(j) > (2/3)T}) \geq (1-w)\deadline$, there is a polynomial time algorithm to place all jobs inside a schedule with height at most $(5/3+\eps)T$.
\end{thm}
\begin{proof}
We place jobs that are very wide or very tall in an ordered fashion, while the residual jobs will be placed using Steinberg's Algorithm, see Figure \ref{fig:Steinberg1}.
We define $\items_{\deadline} := \items_{\width(j) > (1/2 +w)\deadline} \setminus \items_{\height(j) > T/2}$. 
We place each job $j \in \items_{\deadline}$ such that $ \sched(j) = \deadline-\width(j)$.
All the jobs in
$\items_{\height(j) > T/2}$
are sorted by energy demand and placed left aligned, tallest first inside the schedule area.
Let $\rho := \height(\items_{\deadline})/T$ and
let $\height_{(1-2w)\deadline}$ denote the energy demand of the job in $\items_{\height(j) > T/2}$ at position $(1-2w)\deadline$. 
Then $\height_{(1-2w)\deadline} \geq (2/3)T$. 
By Lemma \ref{lem:totalHeightOfWideItems} and the choice of $T$, we know that $\height_{(1-2w)\deadline} + \height(\items_{\deadline}) \leq T \leq \opt$
and hence $\rho \leq (1/3)$.
Let $L$ be a vertical line though the schedule, that is at or strictly left of $(1/2-w)\deadline$ and intersects a job from $\items_{\height(j) > T/2}$ and all the jobs from $\items_{\deadline}$. 
By Lemma \ref{lem:LPackingT4} at and left of $L$ the peak energy demand of the schedule is bounded by $(1+\rho/2)T$. 
On the other hand, right of $L$ the energy demand of the schedule does not increase compared to $L$. 
As a consequence, the peak energy demand in the current schedule is bounded by $(1+\rho/2) T \leq (7/6)T$.
Furthermore, we know that right of $(1-2w)\deadline$ the schedule has a peak energy demand of at most $T$.

Consider the set of jobs $\items_{\height(j)\in((1/3)T,(1/2)T]}$. 
By Lemma \ref{lem:T2} we know  $\width(\items_{\height(j)\in((1/3)T,(1/2)T]}) \leq 2w \cdot \deadline$, since $\items_{\height(j) > (2/3)T} \geq (1-w)\deadline$.
Now we consider two cases.

\noindent{\bf Case A.} If $\eps \leq \rho/2$, 
we place all the jobs in $\items_M := \items_{\height(j)\in((1/3)T,(1/2)T]}$ right-aligned next to each other inside the strip. 
Since they have an energy demand of at most $(1/2)T$ and right of $(1-2w)\deadline$ the schedule has a peak energy demand of at most $T$, the peak energy demand of $(5/3)T$ is not exceeded after adding these jobs.
Define $\lambda := \width(\items_M)/\deadline$. 

Now at each point on the x-axis between $0$ and  $a:=(1-\lambda)\deadline$ the schedule has a height of at most $(1+\rho/2)T$, and, therefore, we can use a height of $b := (2/3-\rho/2 +\eps)T$ to place the residual jobs. 
Let $\items_{res}$ denote the set of residual jobs that still have to be placed.
Note that each job in $\items_{res}$ has a height of at most $(1/3)T$ and a width of at most $(1/2 +w)\deadline$ and the total area of these jobs can be bound by
\begin{align*}
    \area(\items_{res}) &\leq  \deadline T - (2/3)T \cdot (1-w)\deadline - \rho T \cdot (1/2 +w)\deadline - (1/3)T \cdot \lambda \deadline \\
    & = (1/3 + (2/3)w - \rho(1/2 +w) - \lambda/3) \deadline T,
\end{align*}
and hence $2\area(\items_{res}) \leq (2/3 + (4/3)w - \rho(1+2w) - (2/3)\lambda) \deadline T$.
On the other hand, it holds that 
\begin{align*}
& ab - (2\width_{\max} - a)_+(2\height_{\max} - b)_+ \\
& = (2/3-\rho/2+\eps)T (1-\lambda)\deadline - (2(1/2 +w)\deadline - (1-\lambda)\deadline)_+(2(1/3)T - (2/3-\rho/2+\eps)T)_+\\
&= (2/3 -\rho/2 +\eps -(2/3)\lambda +(\rho/2 -\eps)\lambda - (2w +\lambda)_+(\rho/2-\eps)_+)\deadline T\\
&= (2/3 + \eps(1 + 2w) -(1/2+w)\rho -(2/3)\lambda) \deadline T,
\end{align*}
since $\rho/2-\eps \geq 0$.
Hence Steinberg's condition is fulfilled if 
$
(4/3)w-\rho(w+1/2)\leq \eps(1 + 2w)
$, 
which is true since $w \leq (3/4)\eps$.

\noindent{\bf Case B.} On the other hand, if $\rho/2 < \eps$, it holds that $(2/3 +\eps -\rho/2)/2 \geq 1/3$, and  we consider the set $\items_M := \items_{\height(j)\in(((2/3 +\eps -\rho/2)/2)T,(1/2)T]}$, instead of the set $\items_{\height(j)\in((1/3)T,(1/2)T]}$, and place it right-aligned.
Again, we define $\lambda := \width(\items_M)$.
Now, each job in $\items_{res}$ has a height of at most $(1/3 + \eps/2 -\rho/4)T$ and a width of at most $(1/2 +w)\deadline$.
The total area of these jobs can be bounded by 
\begin{align*}
    \area(\items_{res}) \leq &\deadline T - (2/3)T \cdot (1-w)\deadline - \rho T \cdot (1/2 +w)\deadline - (1/3+\eps/2 -\rho/4)T \cdot \lambda \deadline \\
    & = (1/3 + (2/3)w - \rho(1/2 +w) - \lambda(1/3+\eps/2 -\rho/4)) \deadline T,
\end{align*}
and hence $2\area(\items_{res}) \leq (2/3 + 4/3w - \rho(1+2w) - (2/3 +\eps -\rho/2)\lambda) \deadline T$.
On the other hand, it holds that 
\begin{align*}
& ab - (2\width_{\max} -a)_+(2\hmax - b)_+ \\
&= (2/3-\rho/2+\eps)T (1-\lambda)\deadline - (2(1/2 +w)\deadline - (1-\lambda)\deadline)_+(2(1/3+\eps/2 -\rho/4)T \\
& \qquad - (2/3-\rho/2+\eps)T)_+ \\
&=  (2/3+\eps -\rho/2- (2/3-\rho/2+\eps)\lambda) \deadline T,
\end{align*}
Hence, Steinberg's condition is fulfilled if 
$
(4/3-2\rho) w -\rho/2\leq \eps,
$
which is true since $w \leq (3/4)\eps$.

Therefore, in both cases we use Steinberg's algorithm to place the jobs $\items_{\res}$ inside a rectangular container $C$ of height $(2/3 +\eps -\rho)T$ and width $(1-\lambda)\deadline$, which in turn is positioned at $\sched(C) = 0$.
\end{proof}

\subsection{Second Steinberg Case}

\begin{thm}[Second Steinberg Case]
\label{thm:secondStein}
Let $T$ be the lower bound on $\opt$ defined as above.
If $\height(\items_{\width(i) \geq (3/4)\deadline}) > (2/3)T$, then there is a polynomial time algorithm that places all the jobs inside the area $[0,\deadline] \times [0,(5/3)T]$.
\end{thm}
\begin{proof}
In the first step, we place all the jobs in $\items_{\deadline} := \items_{\width(j) > \deadline/2}$ and $\items_{seq} := \items_{\height(j) > T/2}\setminus \items_{\deadline}$ L-shaped.
By Lemma \ref{lem:LPackingT4} the resulting schedule has a peak energy demand of at most $(3/2)\opt$.
Let $\height(\items_{\deadline}) := (2/3 + \rho)T$ and $\width(\items_{seq}) := \lambda \deadline$.
By Lemma \ref{lem:totalHeightOfWideItems2}, we know that, since $\height(\items_{\width(j) \geq (3/4)\deadline}) > (2/3)T$, that $\lambda \deadline \leq 2(\deadline - (3/4)\deadline) = \deadline/2$.

The total amount of work $\area(\items_{res})$ of the residual jobs is bounded by 
$$\area(\items_{res}) \leq \deadline T - (3/4)\deadline\cdot (2/3) T - (1/2)\deadline\cdot \rho T - \lambda \deadline \cdot (1/2)T = (1/2 - \rho/2 -\lambda /2) \deadline T\mathrm{.}$$

On the other hand, there is a rectangular area with width  $a := (1-\lambda )\deadline$ and height $b := ((5/3) - (2/3 +\rho))T = (1-\rho)T \geq (1/2)T$ where we can place the residual jobs.
We will place the residual jobs into this area using Steinberg's algorithm.
This is possible if the Steinberg's condition $2\area(\items_{res}) \leq ab - (2\cdot \width_{\max}(\items_{res}) - a)_+(2\cdot \height_{\max}(\items_{res}) - b)_+$ is fulfilled and each job fits inside the schedule area.
Since $ \width_{\max}(\items_{res}) \leq \deadline/2 \leq a$ and $\height_{\max}(\items_{res}) \leq T/2 < b$, it holds that 
\begin{align*}
&ab - (2\cdot \width_{\max}(\items_{res}) - a)_+(2\cdot \hmax(\items_{res}) - b)_+ \\
&= (1-\lambda )\deadline \cdot (1-\rho)T - (\deadline - (1-\lambda )\deadline)_+(T - (1-\rho )T)_+\\
& = (1-\lambda -\rho)\deadline\cdot T\\
& = 2(1/2 - \rho/2 -\lambda /2) \deadline T
\geq 2\area(\items_{res}).
\end{align*}
The condition is fulfilled, and we can use the free rectangular area to place the residual jobs.
\end{proof}

\section{General algorithm}

	
	

This section inspects schedules generated by the AEPTAS from Theorem \ref{thm:aeptas}, more closely. 
The AEPTAS generates a schedule that fits almost all jobs into an amount of work of peak energy demand  $T \leq (1+\eps)\opt$. 
Left out of the schedule is a set of jobs that has a very small total processing time, where each job can have an energy demand up to $\opt$. 
As such, this set of jobs can be fit into a strip of energy demand $\opt$ and processing time $\gamma \deadline$ for some $\gamma >0$. 
Since we aim to generate a schedule of peak energy demand $(5/3+\eps)\opt$, it does not suffice to simply place this set atop the generated schedule, as this would result in a peak energy demand of $2\opt$. 
Instead, we must find some area in the generated schedule, inside of which we can remove jobs such that an energy demand of $\opt/3$ for a processing time of $\lambda \deadline$ is empty, where $\lambda \in [0,1]$ is a small constant depending on $\eps$. 
Once we have achieved this, and placed the jobs removed by this procedure in a way that does not intersect this strip, we can then place the strip of energy demand $\opt$ at exactly that place, resulting in a schedule of peak energy demand $(5/3+\eps)\opt$. 

If none of the previously mentioned cases (as in Theorem  \ref{thm:firstStein} and \ref{thm:secondStein}, that can be solved using Steinberg's algorithm) apply, then the following Theorem combined with Theorem \ref{thm:aeptas} proves Theorem \ref{53approx}.

\begin{thm}\label{APTASapplication}
Let $\eps \in (0,1/3]$, $\eps ' \leq (3/5)\eps$ and $\gamma \leq (3/40)\eps$. 
Given an instance $\items$ with $\height(\items_{\width(j) > (3/4)\deadline}) \leq (2/3)T'$ and $\width(\items_{\height(j) > (2/3)T'}) \leq (1-(3/4)\eps) \deadline$, for $T' = \max\{T_1,T_2,T_3,T_4\}$ and a schedule $\sched$ (e.g. generated by the APTAS) where almost all jobs are placed such that the peak energy demand is $T \leq (1+\eps')\opt$, and the residual jobs inside an additional box $C_{\gamma}$ of energy demand $T$ and processing time $\gamma \deadline$,
we can find a restructured schedule that places all the jobs up to a schedule with peak energy demand of at most $(5/3+\eps)\opt$.
\end{thm}

\begin{proof}
From the schedule $\sched$, we will generate a new schedule $\sched'$. 
Some jobs will be shifted to new starting positions $\sched'$. 
Other Jobs $j$ that are mentioned in this proof, keep their original starting positions, i.e., $\sched'(j) = \sched(j)$.

If the schedule contains a job $j$ with processing time $\width(j) \in [\gamma,(1-2\gamma)\deadline]$ and energy demand $\height(j) \in [(1/3)T, (2/3)T]$, we proceed as follows: 
Since $\width(j) \leq (1-2\gamma)\deadline$ it holds that $\max \{ \sched(j), \deadline - ( \sched(j) +\width(j))\} \geq \gamma \deadline$.
Let us, w.l.o.g., assume that $ \sched(j) \leq \deadline - ( \sched(j) +\width(j))$, otherwise we mirror the schedule at $\deadline/2$.
We shift the job $j$ completely to the right (by at least $\gamma \deadline$) such that it is positioned at $\sched'(j) := \deadline - \width(j)$.
This increases the peak energy demand to at most $(5/3)T$.
Now the schedule between $ \sched(j)$ and $\sched(j) +\gamma \deadline$ has an energy demand of at most $(2/3)T$.
We place the box $C_{\gamma}$ at $\sched(j)$. 
Since the box has an energy demand of at most $T$, the resulting schedule still has a peak energy demand of at most $(5/3)T \leq (5/3+\eps)\opt$.

\begin{figure}
    \centering
    \begin{tikzpicture}
        \pgfmathsetmacro{\w}{11}
        \pgfmathsetmacro{\h}{2}
        \pgfmathsetmacro{\g}{0.025}
        
        \pgfmathsetmacro{\l}{\w/8}
        \pgfmathsetmacro{\ll}{15*\w/64 -  1*\g*\w/8}
        \pgfmathsetmacro{\lll}{9*\w/32 +  36*\g*\w/32}
        \pgfmathsetmacro{\llll}{3*\w/8+\g*\w/8}
        \draw[gray] (0.5*\w,- 0.00*\h)--(0.5*\w,\h);
        \draw[] (0.5*\w,- 0.05*\h) node[below,black]{$\tau_5$}  --(0.5*\w,0*\h);
        \draw[gray] (\l,- 0.00*\h)  --(\l,\h);
        \draw[] (\l,- 0.05*\h)node[below,black]{$\tau_1$}  --(\l,0*\h);
        \draw[gray] (\ll,- 0.0*\h)  --(\ll,\h);
        \draw[] (\ll,-0.05*\h) node[below,black]{$\tau_2$}  --(\ll,0*\h);
        \draw[gray] (\lll,- 0.0*\h)  --(\lll,\h);
        \draw[] (\lll, -0.05*\h)node[below,black]{$\tau_3$}  -- (\lll,0*\h);
        \draw[gray] (\llll,- 0.0*\h)  --(\llll,\h);
        \draw[] (\llll, -0.05*\h)node[below,black]{$\tau_4$}  --(\llll,0*\h);
        
        \draw[gray] (\w-\l,- 0.00*\h)  --(\w-\l,\h);
        \draw[] (\w-\l,- 0.05*\h)node[below,black]{\tiny{$\deadline{-}\tau_1$}}  --(\w-\l,0*\h);
        \draw[gray] (\w-\ll,- 0.0*\h)  --(\w-\ll,\h);
        \draw[] (\w-\ll,- 0.05*\h)node[below,black]{\tiny{$\deadline{-}\tau_2$}}  --(\w-\ll,0*\h);
        \draw[gray] (\w-\lll,- 0.0*\h)  --(\w-\lll,\h);
        \draw[] (\w-\lll,- 0.05*\h)node[below,black]{\tiny{$\deadline{-}\tau_3$}}  --(\w-\lll,0*\h);
        \draw[gray] (\w-\llll,- 0.0*\h)  --(\w-\llll,\h);
        \draw[] (\w-\llll,- 0.05*\h)node[below,black]{\tiny{$\deadline{-}\tau_4$}}  --(\w-\llll,0*\h);
        
        \draw[dashed] (-0.1*\w,\h) --(1.1*\w,\h) node[right]{$T$};
        
        \node at (\l/2,\h/2) {$S_1$};
        \node at (\ll/2+\l/2,\h/2) {$S_2$};
        \node at (\lll/2+\ll/2,\h/2) {$S_3$};
        \node at (\llll/2+\lll/2,\h/2) {$S_4$};
        \node at (\llll/2 + \w/4,\h/2) {$S_5$};
        \node at (\w-\l/2,\h/2) {$S_1$};
        \node at (\w-\ll/2-\l/2,\h/2) {$S_2$};
        \node at (\w-\lll/2-\ll/2,\h/2) {$S_3$};
        \node at (\w-\llll/2-\lll/2,\h/2) {$S_4$};
        \node at (\w-\llll/2 - \w/4,\h/2) {$S_5$};
        
        \draw (0,\h +0.1*\h) -- (0,0) node[below]{$0$} -- (\w,0) node[below]{$\deadline$}-- (\w,\h +0.1*\h);
        \end{tikzpicture}
    \caption{Splitting the given schedule into segments at time points $\tau_1=\frac{\deadline}{8}$, $\tau_2 = \frac{(15-24\gamma)\deadline}{64}$, $\tau_3 = \frac{(9+11\gamma)\deadline}{32}$, $\tau_4 =\frac{(3+2\gamma)\deadline}{8}$ and $\tau_5 = \frac{\deadline}{2}$.}
    \label{fig:VerticalStrips}
\end{figure}
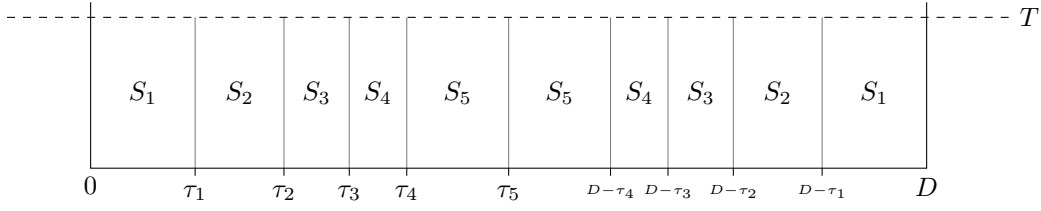

If such a job does not exist, we search for segments that are not overlapped by jobs with energy demand larger than $(2/3)T$.
We split the schedule into segments at the times $\tau_1=\frac{\deadline}{8}$, $\tau_2 = \frac{(15-24\gamma)\deadline}{64}$, $\tau_3 = \frac{(9+11\gamma)\deadline}{32}$, $\tau_4 =\frac{(3+2\gamma)\deadline}{8}$ and $\tau_5 = \frac{\deadline}{2}$ as well at $\tau'_i = \deadline-\tau_i$ for $i \in \{1,2,3,4\}$ and set $\tau_0 = 0$. 
We number the resulting segments in increasing order from $0$ to $\deadline/2$ and from $\deadline$ to $\deadline/2$ such that similar segments on both sides get the same number, see Figure \ref{fig:VerticalStrips}.
We denote by $\sched(S_k)$ ($= \tau_{k-1}$) the start-time of a segment, by $c(S_k)$ ($= \tau_k$) the end-time of the segment and by $\width(S_k)$ ($= \tau_k-\tau_{k-1}$) the processing time of the segment $S_k$ (for $k \in \{1,2,3,4,5\}$). 
Since $\width(\items_{\height(j) > (2/3)}) \leq (1-(3/4)\eps) \deadline$, we know, by pigeonhole principle, that in one of these  segments
a total time of at least $ (3/4)\eps \deadline/10 \geq (3/40)\eps \deadline \geq \gamma \deadline$ is not overlapped by these jobs.

Let $S_{l,k_1}$ be the first such strip from the left, and $S_{r,k_2}$ the last (they might be the same) such that $k_1,k_2 \in \{1,2,3,4,5\}$ represent the index of the strips $S_1,\dots,S_5$. 
In the next step, we modify the start- and end-times of $S_{l,k_1}$ such that it starts at the end of a job with energy demand at least $(2/3)T$ or at $0$.
We denote the shifted start times as $\sched'(\cdot)$ and the shifted completion time as $c'(\cdot)$.
If the start-time of $S_{l,k_1}$, i.e. $\tau_{k-1}$ intersects a job $j$ with $\height(j) \geq (2/3)T$, we define $\sched'(S_{l,k_1}) := \sched(j) +\width(j) \leq c(S_{l,k_1}) -\gamma \deadline $.
Otherwise if $k_1 \not = 1$, we find the last job $j$ ending before $\sched(S_{l,k_1})$ with $\height(j) \geq (2/3)T$ and define  $\sched'(S_{l,k_1}) := \sched(j) + \width(j) \geq \sched(S_{l,k_1}) -\gamma \deadline$ and shift the end-time of $S_{l,k_1}$ by the same amount.
Note that, since $S_{l,k_1}$ is the first strip with at least $\gamma \deadline$ time not occupied by jobs with energy demand larger than $(2/3)T$, the starting time of $S_{l,k_1}$ is reduced by at most $\gamma \deadline$, while the processing time of the segment is not increased.
Finally, if the end-time of $S_{l,k_1}$ intersects a job $j$ that has an energy demand larger than $(2/3)T$, we reduce it to $c'(S_{l,k_1}) := \sched(j)$ and call the modified segment $S'_{l,k_1}$.
These modifications never decreases the total time that is not overlapped by jobs with energy demand larger than $(2/3)T$ in $S_{l,k_1}$.
For an illustration of this procedure, see Figure \ref{fig:shift borders}. 
We do the same but mirrored for $S_{r,k_2}$ resulting in a modified segment $S'_{r,k_2}$. 
If $D -c(S'_{r,k_2}) \leq \sched(S'_{l,k_1})$, we mirror the schedule such that $\sched'(j) = \deadline-c(j)$. 
We denote by $S'_k$ the segment in $\{S'_{l,k_1},S'_{r,k_2}\}$ that appears first in this new schedule, where $k =  \min\{k_1,k_2\}$ represents the original number of the chosen segment.
As a consequence of this mirroring if $k \geq 2$, we ensured there exists a job $j$ with $\height(j) > (2/3)T$ and $c(S'_k) \leq \sched(j) \leq \deadline-\sched(S'_k)$.
Additionally, we know about the start and endpoints of this segment that $\tau_{k-1}-\gamma \deadline \leq \sched(S'_k)$ and $p(S'_k) \leq \tau_k -\tau_{k-1}$.

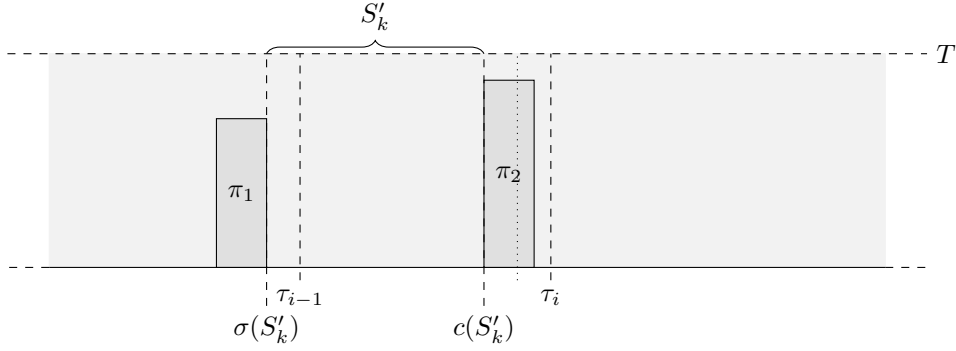
\begin{figure}
    \centering
          \begin{tikzpicture}
            
        \pgfmathsetmacro{\w}{11}
        \pgfmathsetmacro{\h}{1.7}
        \pgfmathsetmacro{\g}{0.025}
        
        \pgfmathsetmacro{\l}{\w/8}
        \pgfmathsetmacro{\ll}{15*\w/64 -  1*\g*\w/8}
        \pgfmathsetmacro{\lll}{9*\w/32 +  36*\g*\w/32}
        \pgfmathsetmacro{\llll}{3*\w/8+\g*\w/8}

        \draw[white!95!black,fill = white!95!black](0*\w, -0.26*\h) rectangle (\w, 5*\h/3-0.26*\h);

        \draw  (0,-0.26*\h) -- (\w,-0.26*\h);
        \draw[dashed]  (0,-0.26*\h) -- (-0.05*\w,-0.26*\h);
        \draw[dashed]  (\w,-0.26*\h) -- (1.05*\w,-0.26*\h);
        
        \draw[dashed] (-0.05*\w,5*\h/3 - 0.26*\h) --(1.05*\w,5*\h/3 - 0.26*\h) node[right]{$T$};
        
        \draw[dashed] (0.3*\w,- 0.36*\h)node[below]{$\tau_{i-1} $}  --(0.3*\w,5*\h/3-0.26*\h);
        \draw[dashed] (0.6*\w,- 0.36*\h)node[below]{$ \tau_{i}$}  --(0.6*\w,5*\h/3-0.26*\h);
        
    	\drawVerticalItem[$\pi_1$]{0.20*\w}{-0.26*\h}{0.26*\w}{\h-0.1*\h};
    	\drawVerticalItem[$\pi_2$]{0.52*\w}{-0.26*\h}{0.58*\w}{\h+0.2*\h};
    	
    	\draw[dashed] (0.26*\w,- 0.56*\h) node[below]{$\sigma(S_k') $} -- (0.26*\w,5*\h/3-0.26*\h);
    	\draw[dashed] (0.52*\w,- 0.56*\h) node[below]{$c(S_k') $} -- (0.52*\w,5*\h/3-0.26*\h);
    	\draw[dotted] (0.56*\w,- 0.36*\h) --(0.56*\w,5*\h/3-0.26*\h);

       \draw [decorate,decoration={brace,amplitude=5pt,raise=0pt},yshift=0pt]
        (0.26*\w,5*\h/3- 0.26*\h) -- (0.52*\w,5*\h/3- 0.26*\h) node [black,midway,yshift=+0.5cm] {
        $S_k'$};

        \end{tikzpicture}
    \caption{An illustration of the border shifting procedure. The original borders are indicated by $\tau_{i-1}$ and $\tau_i$. As $\tau_{i-1}$ is not intersected by a huge job we shift $\sched(S_k')$ to an earlier point in time, such that the huge job $\pi_1$ ends at the exact same time. We then shift $c(S_k')$ by the same amount. 
    The shifted $c(S_k')$ may intersect a huge job $\pi_2$, indicated by the dotted line, and in such a case shift the border further such that  $c(S_k')=\sched(\pi_2)$ holds.}
    \label{fig:shift borders}
\end{figure}

We aim to remove jobs from $S'_k$, such that the peak energy consumption reached inside $S'_k$ is bounded by $(2/3)T$.
We categorize the jobs to be removed in three classes;
first the set of jobs $\itemsBuck$ that are wholly contained in $S'_k$ and have an energy demand less than $(2/3)T$, second the set of jobs that have an energy demand larger than $(2/3)T$, and finally the set of jobs intersecting one of the time points $\sched(S'_k)$ or $c(S'_k)$.

First, we remove $\itemsBuck$ from the segment and schedule them inside a container that has an energy demand of at most $(2/3)T$ and length at most $\width(S'_k)$. 
\begin{lem}
\label{lma:itemsBuck}
The jobs $\itemsBuck$ can be scheduled inside a container $\Cbuck$ of energy demand $(2/3)T$ and processing time $3\width(S'_k) \leq \deadline/2$.
\end{lem}
\begin{proof}
First note that $\area(\itemsBuck) \leq \width(S'_k)T$, since the peak energy demand in $\sched$ is bounded by $T$.
We place these jobs using Steinberg's algorithm. 
Recall that this procedure allows us to place a set of rectangles $\mathcal{R}$ into a container of size $a \cdot b$ as long as the following conditions are met: $\width_{max}(\items)\leq a, \hmax (\items)\leq b,  2 \cdot \area(\items) \leq (ab - (2 \hmax (\items)-T)_+(2\width_{\max}(\items)-\deadline)_+)$. 
Setting our values for $b= 3\width(S'_k)$ and $a=(2/3)T$ yields the desired property. 
Clearly no job wholly contained in a segment of processing time $\width(S'_k)$ can have a processing time greater than $\width(S'_k)$. 
Furthermore, the maximum energy demand of any job in $\itemsBuck$ is $(2/3)T$. 
Finally, we have:
\begin{align*}
    2 \cdot \area(\itemsBuck) & \leq 2\width(S'_k) \cdot T\\
    &\leq 3\width(S'_k)\cdot (2/3)T -(2\width(S'_k) -3\width(S'_k))_+\cdot(2(2/3)T-(2/3)T)_+ \\
    & = (ab - (2 \height_{\max}(\items)-b)_+(2\width_{\max}(\items)-a)_+). \qedhere
\end{align*}
\end{proof}

In the next step, we consider the items with energy demand larger than $(2/3)T$.
By construction of the strip, we know that the total processing time of jobs with energy demands larger than $(2/3)T$ is bounded by $\width(S'_k) -\gamma \deadline$.
We remove all these jobs from the strip and combine them with the extra container $C_{\gamma}$ of energy demand at most $T$ and processing time $\gamma \deadline$ to a new container called $\Ctall$.
It has an energy demand of at most $T$ and a processing time of at most $\width(S'_k)$.

After this step, the only jobs remaining inside the area of $S'_k$ are the jobs that overlap the borders of $S'_k$.
If the peak energy demand in $S'_k$ is lower than $(2/3)T$, we place the container $\Ctall$ inside the strip $S'_k$ as well as  the container $\Cbuck$ right of $\deadline/2$ and are done.
Otherwise, we have to remove jobs that overlap the borders of the strip $S'_k$ until the peak energy demand in $S'_k$ is bounded by $(2/3)T$.
The jobs we choose to remove are dependent on the position of the strip.
The following lemma helps to see how these jobs can be shifted without increasing the peak energy demand of the schedule too much.

\begin{lem}
\label{lma:shiftToRightGA}
Consider a schedule $\sched$ with peak energy demand bounded by $T$ and a time $\tilde{\tau}$, as well as a subset of jobs $\itemsMove \subseteq \items(\tilde{\tau})$ with $e(\itemsMove) \leq a \cdot T$ for some $a \in [0,1]$.
Let $\tau$ be the smallest value $\sigma(j)$ for $j \in \itemsMove$.
Consider the schedule $\sigma'$, where all the jobs in $\itemsMove$ are delayed such that they end at $\deadline$, 
i.\,e., $\sigma'(j) = \deadline-\width(j)$ for all $\itemsMove$ and $\sigma'(j) = \sigma(j)$ for all other jobs.

In the schedule $\sigma'$ before of $\deadline/2+\tau/2$, the peak energy demand is bounded by $T$, while after of $\deadline/2+\tau/2$ the peak energy demand is bounded by $(1+a)T$.
\end{lem}
\begin{proof}
If the energy demand of the schedule $\sigma'$ is larger than $T$ at a position $\tau'$, it has to be because one of the jobs in $\itemsMove$ overlaps it.
Hence, that peak energy demand is bounded by $(1+a)T$, since we shifted jobs with total energy demand bounded by $aT$.
Let $j$ be one of the shifted jobs. 
If $\sched'(j) > \deadline/2+\tau/2$, the energy demand of the schedule before of $\deadline/2+\tau/2$ cannot be influenced by this job.
Therefore, assume that $\deadline- \width(j)  = \sched'(j) \leq \deadline/2+\tau/2$.
As a consequence, $\width(j) \geq \deadline/2-\tau/2$. 
Since $\sched(j) \leq \tau$ it holds that $\sched(j) + \width(j) \geq \deadline/2+ \tau/2$. 
Thus before time $\deadline/2+\tau/2$, the job $j$ overlaps its previous positions and cannot increase the peak energy demand above $T$.
\end{proof}

We choose which of the overlapping jobs to shift depending if $k = 1$ or $k \not = 1$. 
Remember that none of the borders of $S'_k$ overlap a job that has an energy demand larger than $(2/3)T$, and 
assume for the following, that there is a point inside $S'_k$ where the total energy demand of overlapping jobs is larger than $(2/3)T$.

\paragraph*{Case 1: $k = 1$}
Consider the time $\tau = c(S'_1) \leq \deadline/8$ and the set of jobs $\items(\tau)$ that are intersected by this line.
We know that the total energy demand of jobs with processing time greater than $(3/4)\deadline$ is bounded by $(2/3)T$.
Let $\itemsMove$ be the set of jobs generated as follows:
Greedily take the jobs with the largest energy demand from $\items(\tau) \setminus \items_{\width(j) > (3/4)\deadline}$, until either all the jobs from $\items(\tau) \setminus \items_{\width(j) > (3/4)\deadline}$ are contained in $\itemsMove$ or $\height(\itemsMove) \in [(1/3)T, (2/3)T]$.
In this process, we never exceed $(2/3)T$ since, if there is a job with energy demand larger than $(1/3)T$ in $\items(\tau) \setminus \items_{\width(j) > (3/4)\deadline}$, we choose it first and immediately stop.
We delay the jobs in $\itemsMove$ to new start positions $\sched'$ such that for each job $j \in \itemsMove$ we have that $c'(j) := \sched'(j) + \width(j) = \deadline$.
Note that $\sched'(j) \geq (1/4) \deadline$ for each $j \in \itemsMove$ and therefore no longer overlaps $c(S'_1)$.
Furthermore, we know by Lemma \ref{lma:shiftToRightGA} that before $\deadline/2$ the peak energy demand is bounded by $T$, while after $\deadline/2$ the peak energy demand is bounded by $(5/3)T$.
Furthermore, the peak energy demand inside $S'_1$ is bounded by $(2/3)T$.

Since $S'_1$ has a processing time of at most $\deadline/8$, we know by Lemma \ref{lma:itemsBuck} that $\itemsBuck$ can be placed inside a container $\Cbuck$ with energy demand at most $(2/3)T$ and processing time bounded by $3 {\deadline}/{8}$. 
Therefore, we can schedule this container at $\deadline/8$ and know that it is finished before $\deadline/2$. 
Finally, we schedule the container $\Ctall$ at $\sched'(\Ctall) =0$.
The peak energy demand of the resulting schedule is bounded by $(5/3)T$.
See Figure \ref{fig:case k=1} for the repacking procedure.

\begin{figure}
    \centering
    \begin{subfigure}[t]{.32\textwidth}
        \centering
        \resizebox{0.98\textwidth}{!}{

      \begin{tikzpicture}[yscale=0.8]
        \pgfmathsetmacro{\s}{1}
        \pgfmathsetmacro{\w}{6}
        \pgfmathsetmacro{\h}{5}
        \pgfmathsetmacro{\hh}{2/3*\h - 0.26*\h}
        \pgfmathsetmacro{\r}{ 0.26}
        \pgfmathsetmacro{\l}{0.88}
        \draw[white] (-0.1*\w,-0.35*\h- 0.26*\h) rectangle (1.45*\w,5*\h/3 +0.1*\h- 0.26*\h);

        \draw[fill = white!95!black](0*\w, -0.26*\h) rectangle (\w, \h-0.26*\h);
    	
    	\draw[white,fill=white](0.05*\w,2*\h/3-0.3*\h) rectangle (0.1*\w, \h-0.26*\h);
        \draw[white,fill=white](0.02*\w,2*\h/3-0.1*\h) rectangle (0.05*\w, \h-0.26*\h);
        \draw[white,fill=white](0.1*\w,2*\h/3-0.1*\h) rectangle (0.3*\w, \h-0.26*\h);
        \draw[white,fill=white](0.1*\w,2*\h/3-0.3*\h) rectangle (0.2*\w, \h-0.26*\h);

        \drawJobNoBorder{0.9*\w}{4*\h/3-0.4*\h}{\w}{4*\h/3-0.22*\h};
        \drawJobNoBorder{0.8*\w}{\h-0.26*\h}{\w}{4*\h/3-0.4*\h};
         \node at (0.9*\w,\h-0.15*\h) {$\itemsMove$};
         
        \drawVerticalItemRotate[$\Ctall$]{0.0*\w}{2*\h/3-0.3*\h}{0.1*\w}{5*\h/3-0.3*\h};
        \drawVerticalItem[$\Cbuck$]{0.125*\w}{\h-0.26*\h}{0.5*\w}{5*\h/3-0.26*\h};

        \draw[dashed] (-0.05*\w,\h-0.26*\h) --(1.05*\w,\h-0.26*\h) node[right]{$T$};
        \draw[dashed] (-0.05*\w,5*\h/3 - 0.26*\h) --(1.05*\w,5*\h/3 - 0.26*\h) node[right]{$(5/3+\epsilon)T$};
        
        \draw[dashed] (0.0*\w,- 0.26*\h)node[below]{$ $}  --(0.0*\w,\h-0.26*\h);
        \draw[dashed] (0.1*\w,- 0.26*\h)node[below]{$ $}  --(0.1*\w,\h-0.26*\h);
        \draw[dashed] (0.5*\w,- 0.31*\h)node[below]{$\deadline/2$}  --(0.5*\w,\h-0.26*\h);
         \draw [decorate,decoration={brace,amplitude=5pt,mirror,raise=0pt},yshift=0pt]
        (0.0*\w,0*\h- 0.26*\h) -- (0.1*\w,0*\h- 0.26*\h) node [black,midway,yshift=-0.5cm] {
        $S'_1$};
        \draw[dashed] (0.125*\w,- 0.46*\h)node[below]{$\deadline/8$}  --(0.125*\w,\h-0.26*\h);

        \draw (0,5*\h/3 +0.1*\h-0.26*\h) -- (0,-0.26*\h) -- (\w,-0.26*\h)-- (\w,5*\h/3 +0.1*\h-0.26*\h);
        \end{tikzpicture}
    }
    \caption{Repacking for $k=1$}
    \label{fig:case k=1}
    \end{subfigure}
    \begin{subfigure}[t]{.32\textwidth}
        \centering
        \resizebox{0.98\textwidth}{!}{
        \begin{tikzpicture}[yscale=0.8]
            
        \pgfmathsetmacro{\w}{6}
        \pgfmathsetmacro{\h}{5}
        \pgfmathsetmacro{\g}{0.025}
        
        \pgfmathsetmacro{\hh}{2/3*\h - 0.26*\h}
        \pgfmathsetmacro{\r}{ 0.26}
        \pgfmathsetmacro{\ll}{15*\w/64 -  7*\g*\w/8}
        \pgfmathsetmacro{\l}{ 0.88}
        \pgfmathsetmacro{\g}{0.01}
        \draw[white] (-0.1*\w,-0.35*\h- 0.26*\h) rectangle (1.45*\w,5*\h/3 +0.1*\h- 0.26*\h);

        \draw[fill = white!95!black](0*\w, -0.26*\h) rectangle (\w, \h-0.26*\h);
    	
        \draw[white,fill=white](\l/2+\ll/2,2*\h/3-0.1*\h) rectangle (0.4*\w, \h-0.26*\h);
        \draw[white,fill=white](\l/2+\ll/2,2*\h/3-0.3*\h) rectangle (0.3*\w, \h-0.26*\h);

        \drawJobNoBorder{0.9*\w}{4*\h/3-0.4*\h}{\w}{4*\h/3-0.22*\h};
        \drawJobNoBorder{0.8*\w}{\h-0.26*\h}{\w}{4*\h/3-0.4*\h};
        \node at (0.9*\w,\h-0.15*\h) {$\itemsMove$};
         
        \drawVerticalItemRotate[$\Ctall$]{\l-\g}{2*\h/3-0.3*\h}{\ll-\g}{5*\h/3-0.3*\h};
        \drawVerticalItem[$\Cbuck$]{\ll-\g}{\h-0.26*\h}{0.5*\w + \l/2-\g/2}{5*\h/3-0.26*\h};
        
        \drawVerticalItem[$j_r$]{0.6*\w}{-0.26*\h}{0.7*\w}{2*\h/3};
        \drawVerticalItem[$j_v$]{0*\w}{\h-0.26*\h}{0.08*\w}{4*\h/3};
        \draw[dashed] (-0.1*\w,\h-0.26*\h) --(1.1*\w,\h-0.26*\h) node[right]{$T$};
        \draw[dashed] (-0.1*\w,5*\h/3 - 0.26*\h) --(1.1*\w,5*\h/3 - 0.26*\h) node[right]{$(5/3+\epsilon)T$};
       
        \draw[dashed] (0.5*\w,- 0.31*\h) node[below]{$\deadline/2$}  -- (0.5*\w,\h-0.26*\h);
        \draw[dashed] (\l-\g,-0.26*\h) -- (\l-\g,\h-0.26*\h);
        \draw[dashed] (\ll-\g,-0.26*\h) -- (\ll-\g,\h-0.26*\h);
        
        \draw [decorate,decoration={brace,amplitude=5pt,mirror,raise=0pt},yshift=0pt]
        (\l-\g,0*\h- 0.26*\h) -- (\l-\g,0*\h- 0.26*\h) node [black,midway,yshift=-0.4cm] {
        $S'_2$};

        \draw (0,5*\h/3 +0.1*\h-0.26*\h) -- (0,-0.26*\h) -- (\w,-0.26*\h)-- (\w,5*\h/3 +0.1*\h-0.26*\h);
        \end{tikzpicture}
        }
    \caption{Repacking for $k\in \{2,3\}$}
\label{fig:case k=2,3}
    \end{subfigure}
    \begin{subfigure}[t]{.32\textwidth}
        \centering
        \resizebox{0.98\textwidth}{!}{
      \begin{tikzpicture}[yscale=0.8]
            
        \pgfmathsetmacro{\w}{6}
        \pgfmathsetmacro{\h}{5}
        \pgfmathsetmacro{\hh}{2/3*\h - 0.26*\h}
        \pgfmathsetmacro{\g}{0.025}
        \pgfmathsetmacro{\r}{ 0.26}
        \pgfmathsetmacro{\l}{ 0.88}
        \pgfmathsetmacro{\ll}{15*\w/64 -  1*\g*\w/8}
        \pgfmathsetmacro{\lll}{9*\w/32 +  36*\g*\w/32}
         \pgfmathsetmacro{\llll}{3*\w/8+\g*\w/8}
        \draw[white] (-0.1*\w,-0.35*\h- 0.26*\h) rectangle (1.45*\w,5*\h/3 +0.1*\h- 0.26*\h);

        \draw[fill = white!95!black](0*\w, -0.26*\h) rectangle (\w, \h-0.26*\h);

        \draw[white,fill=white](\lll,2*\h/3-0.1*\h) rectangle (\w/2, \h-0.26*\h);
        \draw[white,fill=white](\lll,2*\h/3-0.3*\h) rectangle (\llll+0.05*\w, \h-0.26*\h);

        \drawJobNoBorder{\w-\llll-0.05*\w+\lll}{4*\h/3-0.4*\h}{\w}{4*\h/3-0.22*\h};
        \drawJobNoBorder{\w/2+\lll}{\h-0.26*\h}{\w}{4*\h/3-0.4*\h};
         \node at (0.9*\w,\h-0.15*\h) {$\itemsMove$};

        \drawVerticalItemRotate[$\Ctall$]{\lll}{2*\h/3-0.3*\h}{\llll}{5*\h/3-0.3*\h};
  
        \drawVerticalItem[$\Cbuck$]{0}{\h-0.26*\h}{\ll}{5*\h/3-0.26*\h};
        
        \drawVerticalItem[$j_v$]{\llll}{\h-0.26*\h}{\llll+0.08*\w}{5*\h/3-0.26*\h};
        
        \drawVerticalItem[$j_r$]{0.56*\w}{-0.26*\h}{0.63*\w}{2*\h/3-0.05*\h};
        
        \draw (0,5*\h/3 +0.1*\h-0.26*\h) -- (0,-0.26*\h) -- (\w,-0.26*\h)-- (\w,5*\h/3 +0.1*\h-0.26*\h);

        \draw[dashed] (-0.1*\w,\h-0.26*\h) -- (1.1*\w,\h-0.26*\h) node[right]{$T$};

        \draw[dashed] (-0.1*\w,5*\h/3 - 0.26*\h) -- (1.1*\w,5*\h/3 - 0.26*\h) node[right]{$(5/3+\epsilon)T$};

        \draw[dashed] (0.5*\w,- 0.31*\h)node[below]{$\deadline/2$}  --(0.5*\w,\h-0.26*\h);
         \draw [decorate,decoration={brace,amplitude=5pt,mirror,raise=0pt},yshift=0pt]
        (\lll,0*\h- 0.26*\h) -- (\llll,0*\h- 0.26*\h) node [black,midway,yshift=-0.4cm] {
        $S'_4$};
      
         \draw[dashed] (\lll,-0.26*\h) -- (\lll,\h-0.26*\h);
         \draw[dashed] (\llll,-0.26*\h) -- (\llll,\h-0.26*\h);
 
        \end{tikzpicture}
        }
    \caption{Repacking for $k\in \{4,5\}$}
    \label{fig:case k =4,5}
    \end{subfigure}
    \caption{Illustration of the steps in the proof of \Cref{APTASapplication}. Note that the set $\itemsMove$ is delayed such that the jobs end at $\deadline$. The containers $\Ctall$ and $\Cbuck$ are placed such that they do not intersect. For \ref{fig:case k=2,3} and \ref{fig:case k =4,5}, the jobs $j_v$ are placed in the same manner, and the job $j_r$ is denoted. }
    \label{fig:repackingGrouped}
\end{figure}

\paragraph*{Case 2: $k \not = 1$}
In this case, the borders of the considered strip can be overlapped from both sides.
Furthermore, we know that the left border of $S'_k$ is right of $\deadline/8-\gamma \deadline\geq \gamma \deadline$.

Consider the largest total energy demand of jobs that are intersected by any vertical line through $S'_k$ and denote this energy demand as $T_{S'_k}$.
Since there is a job $j_l$ with energy demand larger than $(2/3)T$ with $ \sched(j_l) +\width(j_l) = \sched(S'_k)$, we know that the total energy demand of jobs intersecting $\sched(S'_k)$ can be at most $(1/3)T$. 
Next, consider the closest job $j_r$ that starts after $c(S'_k)$ and has an energy demand larger than $(2/3)T$. 
By the choice of $S'_k$, we know that such a job must exist and that $\sched(j_r) \leq \deadline - \sched(S'_k)$, 
by the choice of $S'_k$ out of $S'_{k_1,l}$ and $S'_{k_2,r}$.
Furthermore, we know that 
the total energy demand of jobs intersecting the vertical line at $\sched(j_r)$ is bounded by $(1/3)T$.

Hence the jobs that overlap the vertical line at $\sched(j_r)$ and the jobs that overlap the vertical line at $\sched(S'_k)$ add a total energy demand of at most $(2/3)T$ to $T_{S'_k}$.
Let us now consider the jobs $\items_M$ that overlap the time $c(S'_k)$ but neither the time $\sched(S'_k)$ nor the time $\sched(j_r)$. 
Each of them has a processing time of at most $\deadline - \sched(S'_k) - \sched(j_r) \leq \deadline-2 \sched(S'_k)$.
Hence when delaying their start points such that $\sched'(j) = \deadline-p(j)$, they no longer overlap the time $c(S'_k)$ since $\width(S'_k) \leq \sched(S'_k)$ for each $k \in \{2,3,4,5\}$.

We greedily take jobs from $\items_M$ that have the earliest starting point until we have all jobs from $\items_M$ or we have a total energy demand of at least $(1/3)T$.
If the total energy demand of the chosen jobs is larger than $(2/3)T$, the last job $j_v$ has an energy demand of at least $(1/3)T$.
Since it has a processing time lower than $(1-2\gamma)\deadline$, it has to have a processing time of at most $\gamma\deadline$.
We remove this job and place it later, while we shift all the others to new positions $\sched'$ such that $\sched'(j) = \deadline-\width(j)$ for each of the taken jobs $j$.
We call the set of shifted jobs $\itemsMove$.
Furthermore, since $\itemsMove$ has a total energy demand of at most $(2/3)T$ and a starting point right of $\sched(S'_k)$, we know by Lemma \ref{lma:shiftToRightGA} that the peak energy demand right of $\deadline/2 + \sched(S'_k)/2$ is bounded by $(5/3)T$ while left of $\deadline/2 + \sched(S'_k)/2$ it is bounded by $T$.

Let $\sched(j_l)$ be the starting time of the last taken job.
Before $\sched(j_l)$ (in $S'_k$) there is no longer a job from $\items_M$, and, hence inside $S'_i$ left of $\sched(j_l)$, the peak energy demand is bounded by  $(2/3)T$.
On the other hand, after $\sched(j_l)$ (in $S'_k$) we either have removed jobs with total energy demand at least $(1/3)T$, or all the jobs from $\items_M$ and hence the schedule there can have a total energy demand of at most $(2/3)T$ as well.
Therefore, we can place the container $\Ctall$ inside $S'_k$ without increasing the energy demand above $(5/3)T$.

It remains to place the container $\Cbuck$ and the job $j_v$.
For $k \in\{2,3\}$, we set  $\Cbuck$ at $\sched'(\Cbuck) = c(S'_k)$ and the job $\sched'(j_v) = 0$,
while for $k \in \{4,5\}$, we set $\sched'(\Cbuck) = 0$ and $\sched'(j_v) = c(S'_k)$.
We will now see for each segment, that the peak energy demand of $(5/3)T$ is not exceeded by this new schedule.

First note that $p(j_v) \leq \gamma \deadline \leq \deadline/8 -\gamma\deadline$ and hence does not intersect $S'_2$, when scheduled at $\sigma'(j_v) = 0$. Similarly, it is more narrow than $S'_5$ and $\sched(S'_5)/2$, and hence fits right of $S'_4$ and $S'_5$ without increasing the schedule more than $(5/3)T$.

It remains to place the container $\Cbuck$ and the job $j_v$.
For $k \in\{2,3\}$, we set  $\Cbuck$ at $\sched'(\Cbuck) = c(S'_k)$ and the job $\sched'(j_v) = 0$,
while for $k \in \{4,5\}$, we set $\sched'(\Cbuck) = 0$ and $\sched'(j_v) = c(S'_k)$.
We will now see for each segment, that the peak energy demand of $(5/3)T$ is not exceeded by this new schedule.


Let us now check the conditions for $\Cbuck$:
For $k \in\{2,3\}$, we have to ensure that $c(S'_k) + \width(\Cbuck)/2 \leq \deadline/2 + \sched(S'_k)$, while for $k \in \{4,5\}$ we have to prove that $p(\Cbuck) \leq \sched(S'_k)$.
It holds that $\width(S'_2) \leq \frac{(15-24\gamma)\deadline}{64}-\frac{\deadline}{8} = \frac{(7-24\gamma)\deadline}{64}$ and hence $p(\Cbuck) \leq 3(\frac{(7-24\gamma)\deadline}{64})$. 
Therefore,  $$c(S'_2) + p(\Cbuck) \leq \frac{(15-24\gamma)\deadline}{64} + 3\left(\frac{(7-24\gamma)\deadline}{64}\right) \leq \frac{\deadline}{2} + \frac{\deadline/8 - \gamma \deadline}{2} \leq \frac{\deadline}{2} + \frac{\sched(S'_2)}{2}.$$ 
Furthermore, 
$$p(S'_3) \leq \frac{(9+14\gamma)\deadline}{32}-\frac{(15-24\gamma)\deadline}{64} = (\frac{3}{64}+\frac{52 \gamma}{32})\deadline$$ 
and hence $p(\Cbuck) \leq 3(\frac{3}{64}+\frac{52 \gamma}{32})\deadline$.
Therefore, $$c(S'_3) + p(\Cbuck) \leq \frac{(9+14\gamma)\deadline}{32} + 3(\frac{3}{64}+\frac{52 \gamma}{32})\deadline = \frac{(27 + 184\gamma)\deadline}{64},$$
while $\deadline/2 + \sched(S'_3)/2 \geq \deadline/2 + (\frac{(15-24\gamma)}{64} - \gamma)\deadline/2 = (\frac{79}{128} - \frac{11}{8}\gamma)\deadline$.
As a consequence, $c(S'_3) + p(\Cbuck) \leq \deadline/2 + \sched(S'_3)/2$, since $\gamma\leq 1/40 \leq 25/392$.

Finally, we have 
$p(S'_4) \leq \frac{(3+2\gamma)\deadline}{8}-\frac{(9+14\gamma)\deadline}{32} = \frac{(3-6\gamma)\deadline}{32}$.  
Hence,  
$$p(\Cbuck) \leq 3\left(\frac{(3-6\gamma)\deadline}{32}\right) =  \frac{(9+14\gamma)\deadline}{32}   -\gamma \deadline \leq \sched(S'_4).$$ 
While it holds that 
$p(S'_5) \leq \frac{\deadline}{2} - \frac{(3+2\gamma)\deadline}{8} = \frac{(1+2\gamma)\deadline}{8}$.  
Therefore, 
$$p(\Cbuck)\leq 3\cdot\frac{(1+2\gamma)\deadline}{8} = \frac{(3+6\gamma)\deadline}{8} = \frac{(1+2\gamma)\deadline}{8} -\gamma \deadline \leq  \sched(S'_5).$$

For a visual representation of this repacking procedure see Figure \ref{fig:repackingGrouped}. In all the cases the generated schedule has a height of at most $(5/3)T \leq (5/3)(1+\eps')\opt \leq (5/3 +\eps)\opt$.

\end{proof}

\section{AEPTAS for NPDM}

In this section, we will prove the following theorem.
\begin{thm}
\label{thm:aeptas}
Let $\eps > 0,\lambda \in \Oh(\eps)$. 
There is an algorithm that places almost all jobs such that the peak energy demand is bounded by $T' := (1+\Oh(\eps))\opt$ and all residual jobs are placed either in a container $C_1$ with processing time $\lambda \deadline$ and energy demand $T'$ or in a container $C_2$ with processing time  $\deadline$ and energy demand $\hmax$. 
The time complexity of this algorithm is bounded by $\Oh(n \log(n)/\eps) + 1/\eps^{1/\eps^{\Oh(1/\eps)}}$.
\end{thm}

The statement gives in fact two variants of the 
algorithm. The first variant where all residual jobs are placed in $C_1$ is used in our $5/3 +\eps$
approximation algorithm, where the second variant with all residual jobs in $C_2$ can be used to obtain the AEPTAS.
The described main algorithm follows the dual-approximation framework.
We describe an algorithm that given a bound on the schedule peak energy demand $T$ computes a schedule with peak energy demand $(1+\Oh(\eps))T' + \hmax$ or decides correctly that there is no schedule with peak energy demand at most $T'$.
This algorithm then can be called in binary search fashion with values $T$ between $T' = \max\{T_1,T_2,T_3,T_4\}$  and $\max\{2 \area(\items)/\deadline,2\hmax\}$, using only multiples of $\eps T'$.

Note that if $\hmax \leq \mathcal{O}(\eps^3T')$, we can use the algorithm in \cite{DBLP:journals/dmaa/BougeretDJRT11} 
to find an $(1+\eps)\opt + \mathcal{O}(\log(1/\eps)/\eps \cdot \eps^3T') = (1+\mathcal{O}(\eps))\opt$ approximation.  
Hence we can assume that $\hmax > \mathcal{O}(\eps^3T')$. 

\paragraph*{Classification of Jobs}

Given two values $\delta$ and $\mu$ with $\mu < \delta$, we partition the jobs into five sets: large, horizontal, vertical, small, and medium sized jobs.
\begin{itemize}
\item $\Rla:= \{i \in \jobs| \height(i) \geq \delta T', \width(i) > \delta \deadline\}$
\item $\Rho:= \{i \in \jobs| \height(i) < \mu T', \width(i) > \delta \deadline\}$
\item $\Rve:= \{i \in \jobs| \height(i) \geq \delta T', \width(i) < \mu \deadline\}$
\item $\Rsm:= \{i \in \jobs| \height(i) < \mu T', \width(i) < \mu \deadline\}$
\item $\jobs_{medium}:= \jobs \setminus (\Rla \cup \Rho \cup \Rve \cup \Rsm)$
\end{itemize}

\begin{lem}
    In $\Oh(n + 1/\eps^2)$ operations it is possible to find values $\geq \eps^{\mathcal{O}(1/\eps^2)}$ for $\delta$ and $\mu$ such that $\area(\jobs_{medium}) \leq  (\eps^2/4) \deadline T$ and $\mu \leq  c\eps^5 \delta$ for any given constant $c$.
\end{lem}
\begin{proof}
    Consider the sequence $\rho_0 := \eps^5/4$, $\rho_{i+1} := c\rho_{i} \eps^3$. 
    Due to the pigeonhole principle, there exists an $i \in \{0,\dots,8/\eps^2\}$ such that when defining $\delta := \sigma_i$ and $\mu := \sigma_{i+1}$ the total amount of work of the medium sized jobs is bounded by $(\eps^2/4) \deadline T$, because each job appears only in two possible sets of medium jobs. 
    We have $\delta \geq \mu \geq \eps^{\mathcal{O}(1/\eps^2)}$.
\end{proof}


\begin{lem} \label{geo:round} \cite{rau2019useful}
We can round the energy demands $e(i)$ of the vertical and large jobs to multiples $k_i \eps \delta T$ with $k_i \in \{1/\eps,\ldots,1/\eps\delta\}$ such that the number of different demands is bounded by $O(1/\eps^2 \log(1/\delta))$.
\end{lem}

This can be done via a standard combination of arithmetic and geometric rounding \cite{rau2019useful}. 
In the following we will dismiss the medium jobs from the schedule.


\subsubsection*{Profile for vertical jobs} 
Given an optimal schedule, we partition the schedule into $1/\gamma$ segments of processing time $\gamma \deadline$, for a constant $\gamma \in \mathcal{O}_{\eps}(1)$.
Given a schedule of jobs $J$, we define profile of $J$ to be $\{(x,y)|y=\sum_{j \in J| \sigma(j)\le x\le \sigma(j)+\width(j)} \height(j), 0 \le x\le D, \}$.
Energy demand of profile of jobs $J$ at time $t$ is $\mathcal{E}_J(t):=\sum_{j \in J| \sigma(j)\le t \le \sigma(j)+\width(j)} \height(j)$. 
Now consider the profile of large and horizontal jobs.
Let  $\tilde{J}:=\Rla \cup \Rho$,
We search for the segments where the maximal energy demand of the profile of large and horizontal jobs and the minimal energy demand of this profile differs more than $\eps T$, i.e., if in segment $S:=(t_a, t_b)$, $|\max_{t \in S} \mathcal{E}_{\tilde{J}}(t) - 
\min_{t \in S} \mathcal{E}_{\tilde{J}}(t)| \ge \eps T$,  then we remove all vertical and small jobs from these segments fractionally, i.e., we slice jobs, which are cut by the borders of the segment. 




\begin{claim}
Let $\jobs_{rem}$ be the set of removed vertical and small jobs. 
Then $\area(\jobs_{rem})$ is bounded by $\mathcal{O}(\gamma/\eps\delta) \cdot \deadline \cdot T$.
\end{claim}

\begin{proof}
Note that the energy demand of the profile of horizontal or large jobs only changes, when horizontal or large jobs end or start. 
The large and horizontal jobs have a total energy demand of at most $T/\delta$ since they have a processing time of at least $\delta \deadline$ and the total area of the schedule is bounded by $T \cdot \deadline$. 
Hence there can be at most $2 (T /\delta) /\eps T = \mathcal{O}(1/\eps\delta)$ segments, where the energy demand of the profile changes more than $\eps T$.
As a result, the total area of the removed vertical jobs can be bounded by $\mathcal{O}(1/\eps\delta) \cdot (\gamma \deadline \cdot T)$.
\end{proof}

%

\begin{claim}[Size of $\gamma$]
In the case of container $C_1$, we can choose 
$\gamma \in \Oh(\eps\delta\lambda)$ such that we 
can schedule the removed vertical jobs fractionally inside a container $C_{1,1/4}$ of processing time $\width(C_{1})/4$ and energy demand $\height(C_{1})$.
Otherwise, we can choose $\gamma \in \Oh(\eps^4\delta)$ such that we 
can schedule the removed vertical jobs fractionally inside a container $C_{2,1/4}$ of processing time $p(C_{2})/4$ and energy demand $\height(C_{2})$.
\end{claim}

\begin{proof}
Let $k \in \{1,2\}$ depending on the chosen container.
First we place all the jobs $\jobs_{rem,tall}$, i.e., jobs in $\jobs_{rem}$ with energy demand larger than $\height(C_{k,1/4})/2$ next to each other.
The total processing time of these jobs is bounded by $2\cdot \area(\jobs_{rem,tall})/\height(C_{k,1/4})$.
Next, we place the residual jobs $\jobs_{rem,res}:=\jobs_{rem}\setminus \jobs_{rem,tall}$, which have an energy demand of at most $\height(C_{k,1/4})/2$.
We take slices of processing time $1$ of the jobs and place them on top of each other until the energy demand $\height(C_{k,1/4})/2$ is reached.
Since each job has an energy demand of at most $\height(C_{k,1/4})/2$ the energy demand $\height(C_{k,1/4})$ is not exceeded. 
The total processing time of this schedule is bounded by
$2 \area(\jobs_{rem,res})/\height(C_{k,1/4}) + 1 \leq \Oh(\gamma/(\eps\delta) \cdot \deadline \cdot T)/\height(C_{k,1/4})$.
Hence, for $C_{1,1/4}$
the total processing time is bounded by $\Oh(\gamma/(\eps\delta) \cdot \deadline \cdot T)/ T = \Oh(\gamma/(\eps\delta)) \deadline$.
Hence, when choosing $\gamma \in \Oh(\lambda \eps \delta)$ for a suitable constant, the total processing time of this schedule is bounded by $p(C_{1})/4$. 
Otherwise, for container $C_{2,1/4}$ the total processing time is bounded by $\Oh ((\gamma/(\eps\delta) \cdot \deadline \cdot T)/ \eps^3T) = \Oh(\gamma/\eps^4\delta) \deadline$.
Hence, when choosing $\gamma \in \Oh(\eps^4\delta)$ for a suitable constant, the total processing time of this schedule is bounded by $p(C_{2})/4$. 
\end{proof}

\paragraph*{Algorithm to place the vertical, small, and medium jobs}
In the algorithm, we first round the energy demands of the vertical jobs to at most $\mathcal{O}(1/\eps^2 \cdot \log(1/\delta)) = (1/\eps)^{\Oh(1)}$ sizes using Lemma \ref{geo:round} (geometric rounding).

Afterward, we guess for each of the $1/\gamma$ segments the energy demand reserved for the vertical and small jobs rounding up to the next multiple of $\eps T$, adding at most one more $\eps T$ to the energy demand of the schedule. 
There are at most $\Oh((1/\eps)^{1/\gamma})$ possible guesses.
Furthermore, we introduce one segment $\top$ of energy demand $\lceil \height(C_{k})/(\eps T) \rceil \cdot {\eps T}$ and processing time $\width(C_{k})/4$ ($k \in \{1,2\}$) for the set of removed vertical jobs.
Let $S_{ver}$ be the set of all introduced segments, and for each $s \in S_{ver}$ let $\height_{s,ver}$ be the energy demand reserved for vertical and small jobs.
Furthermore, let $S_{ver,e}$ be the set of segments that have exactly energy demand $e$ and let $\width(S_{ver,e})$ be their total processing time.

To place the vertical jobs into the segments $S_{ver}$, we use a configuration LP. 
Let $C = \{a_{\height_i}:\height_i| \height_i \in \{\height(j)| j\in\Rve\}\}$ be a configuration for vertical jobs, $\height(C) := \sum_{i \in \Rve}a_i\height(i)$ be its energy demand, and $\mathcal{C}_{e}$ be the set of configurations with energy demand at most $e$. 
Furthermore, for a given configuration $C$ we denote by $a_{\height}(C)$ the number of jobs contained in $C$ that have an energy demand of $\height$.
Since each vertical job has a energy demand of at least $\delta T$, there are at most $(1/\eps)^{\Oh(1/\delta)}$ different configurations.
Consider the following linear program:
\begin{align}
\sum_{C \in \mathcal{C}_{i \eps T}} x_{C,i\eps T} & = \width(S_{ver,i\eps T}) & \forall i \in \{1,\dots,1/\eps\} \\
\sum_{s \in S}\sum_{C \in \mathcal{C}_{\height_{s,ver}}} a_{\height_i}(C)x_{C,s} &= \sum_{j \in I,\height(j) =\height_i}\width(j) & \forall \height_i \in \{\height(j)| j\in\Rve\}\\
x_{C,s} &\geq 0 & \forall C \in \mathcal{C}, s \in S_{ver}
\end{align}
The first equation ensures that the configurations and the jobs placed on top of the schedule do not exceed the available space, while the second equation ensures that each job is fully scheduled. 
A basic solution has at most $(1/\eps+|\{\height(j)| j\in\Rve\}|+1) = (1/\eps)^{3}$ nonzero components. 

We can solve the above linear program by guessing the set of non zero components and then solving the resulting LP in $((1/\eps)^{\Oh(1/\delta)})^{(1/\eps)^{3}}$ time.

To place the vertical jobs, we  first fill them greedily inside the configurations (slicing when the corresponding configuration slot is full) and afterward place the configurations inside the schedule, slicing the jobs at the segment borders.
For each nonzero component we have one configuration that contains at most $2/\delta$ fractional placed vertical jobs on top of each other, which have a total energy demand of at most $2T'$. 
Additionally, for each segment we have the same amount of fractional jobs.
Hence total area of fractionally placed jobs can be bounded by 
$\mu \deadline \cdot 2T \cdot ((1/\eps)^{3} + 1/\gamma)$.
If we choose $C_1$ this can be bounded by $\Oh(\mu /(\lambda \eps \delta))\deadline T\leq \lambda \deadline T/8$, since $\mu = c \eps \delta \lambda^2$ and otherwise by 
$\in \Oh(\mu \deadline \cdot T/(\eps^5 \delta)) \leq \deadline T/8$, 
since $\mu = c \eps^5 \delta$ 
for a suitable small constant $c$. 
We remove the fractionally placed jobs $\jobs_{frac}$. 



Next, we place the small jobs inside the empty area that can appear above each configuration for vertical jobs.
Note that there are at most  $((1/\eps)^{\Oh(1)} + 1/\gamma)$ configurations and the free area inside these configurations has at least the size of the total area of the small jobs.
As a consequence, we have at most $2/\gamma$ rectangular areas to place the small jobs, which have a total area, which is at least the size of the small jobs.
We use the NFDH algorithm to place these jobs inside the boxes until no other job fits inside.

Assume we could not place all the small jobs inside these boxes. 
When considering the free amount of work in each box, there are three parts which contribute to it. 
First, each box can have a strip of free amount of work later on in the schedule which has a processing time of at most $\mu \deadline$.
The total free amount of work contributed by this strip is bounded by $(2/\gamma) \mu \deadline \cdot 2T$.
Second, each box can have a strip of free amount of work of energy demand at most $\mu T$ on the top, because otherwise another line of jobs would have fitted inside this box. 
Since there are no boxes on top of each other, we can bound the total free amount of work which is a result of this strip by $\mu T \cdot \deadline$.
Finally, there can be free amount of work between the shelves of the jobs generated by the NFDH algorithm.
This total free amount of work is bounded by the energy demand of the tallest job times the processing time of the widest box, i.e. $\mu T \cdot \gamma \deadline$.
Hence the total free amount of work inside the boxes is bounded by $5 \mu \deadline \cdot T \cdot \gamma + \deadline \cdot \mu T$.
Since $\gamma \in \Oh(1/\eps^5\delta)$ and we have chosen $\mu \leq c \delta \eps^6 $ for a suitable small constant $c \in \mathbb{Q}$, the total amount of work of the residual small jobs $\jobs_{small, res}$, which could not be placed is bounded by $\eps T \deadline$.

We place the residual small jobs $\jobs_{small, res}$ on top of the schedule using NFDH.
This adds an energy demand of at most $2\eps T$ to the schedule.
Next we place the medium jobs.
We start all the medium jobs, that have a processing time larger than $\width(C_{k})/4$ with Steinberg's algorithm inside a box of energy demand at most $\Oh(\eps) T$ and processing time $\deadline$. 
This is possible since they have processing time in $(\eps \deadline,\deadline]$ and therefore each has an energy demand of at most $\Oh(\eps) T$ because their total work is bounded by $(\eps^2/4) \deadline T$.
The residual jobs (that might have a processing time larger than $\eps T$) are placed inside the first half of the container using Steinberg's algorithm. 
The later half of the container is filled with the extra box for vertical items defined for the LP and the fractionally scheduled jobs. 
The extra box has a width of at most $\width(C_{k})/4$. 
Since the fractionally placed vertical jobs $\jobs_{frac}$ have an area of at most $\height(C_{k}) \cdot \width(C_{k})/8$ and each has a width of at most $\mu \deadline < \width(C_{k})/8$, we can use Steinberg's algorithm to place them inside the last quarter of the container $C_{k}$.

\subsubsection*{Placement of horizontal jobs}
In this section, we first reduce the number of possible starting points for horizontal jobs and then use a linear program to place the jobs in the schedule.

First step: use geometric grouping to reduce the number of processing times of horizontal jobs.
At a loss of at most $2\eps T$ in the approximation ratio, we can reduce the number of processing times of horizontal jobs to $\mathcal{O}(\log(1/\delta)/\eps)$ 
using geometric grouping (see \cite[Theorem 2]{DBLP:conf/focs/KarmarkarK82} by Karmarkar and Karp).
These rounded jobs can be placed fractionally instead of the original jobs and an extra box of energy demand at most $\Oh(\eps)T$.
Remember, we know the profile of large and horizontal jobs with precision $\eps T$ for the segments of processing time $\gamma \deadline$.  
In the next step, we will alter the starting points of the large and fractionally placed horizontal jobs without exceeding the given profile.

\begin{claim}
Without loss in the approximation ratio, we can reduce the number of different starting points of rounded horizontal and large jobs to 
$ (1/\eps)^{(1/\eps)^{\Oh(1/\eps)}}$.
\end{claim}
\begin{proof}
Consider the large and horizontal jobs starting in the first segment. 
Since this segment has a processing time of $\gamma \deadline \leq \delta \deadline$, there can be no job ending in this segment. 
Hence this segment is maximally filled at the point $\gamma \deadline$. 
We can shift the start point of each job in this segment to $0$ and we will not change the maximal energy demand of this segment.

Now consider a job $i \in \Rho\cup \Rla$ starting in the second segment. 
If there is no horizontal or large job ending before the start of $i$, we can shift the start point of $i$ to $\gamma \deadline$ without changing the maximal filling energy demand in this segment.
However, if there is a job $j \in \Rho\cup \Rla$ ending before $i$ in this segment, we can not shift this job to $\gamma \deadline$ since then $i$ and $j$ overlap, which they did not before.  
This could change the maximal energy demand of the profile in this segment.
Nevertheless, if $j$ is the last job ending before $i$, we can shift $i$ to the left, such that $i$ starts at the endpoint of $j$.

We iterate this shifting with all segments and all jobs in $\Rho\cup \Rla$. 
As a result, all jobs start either at a multiple of $\gamma \deadline$, or they start at an endpoint of an other job in $\Rho\cup \Rla$.
Therefore, we can describe the set of possible starting points for jobs in $\Rho\cup \Rla$ as $S_{hor,large} := \{l\gamma \deadline + \sum_{j = 1}^{1/ \delta} \width(i_j)| l \in \{0,1,\dots,1/\gamma\}, i_j \in \Rho\cup \Rla \forall j \in \{1,\dots,1/\delta_w\}\}$.
It holds that $|S_{hor,large}| \leq (1/\gamma) \cdot (\log(1/\delta)/\epsilon)^{1/\delta} =  (1/\eps)^{(1/\eps)^{\Oh(1/\eps)}}$.
\end{proof}

\begin{claim}
\label{clm:StartingPintsHorizontalImproved}
At a loss of at most $\mathcal{O}(\eps T)$ in the approximation ratio, we can reduce the number of \emph{used} starting points for rounded horizontal jobs to $\mathcal{O}(1/\eps\delta)$.
\end{claim}

\begin{proof}
We partition the set of horizontal jobs by their processing time into $\mathcal{O}(\log(1/\delta))$ sets $\Rho^{l} := \{i \in \Rho| \deadline/2^l < \width(i) \leq \deadline/2^{l-1}\}$.
For each of these sets, we will reduce the number of starting positions to $2^l/\eps^2$. 
We partition the schedule into $2^l$ segments of processing time $\deadline/2^l$. Each job from the set $\Rho^{l}$ has a processing time larger than $\deadline/2^l$ and hence it starts in an other segment as it ends.
We consider for each segment all the horizontal jobs of the set $\Rho^{l}$ ending in this segment and sort them by increasing starting position. 
Let $\height_{l,i}$ be the energy demand of the stack of jobs in $\Rho^{l}$ ending in the $i$-th segment.
We partition the stack into $1/\eps$ layers of energy demand $\eps \height_{l,i}$ and slice the horizontal jobs overlapping the layer borders.  
We remove all the jobs in the bottom most layer and shift the jobs from the layers above to the left, such that they start at the latest original start position from the layer below. 
We repeat this procedure for each segment.
By this shift, we reduce the total number of starting positions from jobs from the set $\Rho^{l}$ to $2^l/\eps$.
The total energy demand of the jobs we removed is bounded by $\eps \height(\Rho^{l})$. 
Since these jobs have a processing time of at most $\deadline/2^{l-1}$, we can schedule $2^{l-1}$ of these jobs after one an other (horizontally), without violating the deadline. 
Hence, when scheduling these jobs fractionally, we add at most $\eps \height(\Rho^{l})/2^{l-1}$ to the schedule. 
Note that since all the jobs in set $\Rho^{l}$ have a processing time of at least $\deadline/2^l$, it holds that $\sum_{l = 1}^{\lceil\log(1/\delta)\rceil} \height(\Rho^{l})/2^l \leq T$ and, hence, we add at most $\sum_{l = 1}^{\lceil\log(1/\delta)\rceil}\eps\height(\Rho^{l})/2^{l-1} \leq 2\eps T$ to the energy demand of the schedule, when scheduling the removed horizontal jobs.

The total number of starting positions is bounded by $\sum_{l = 1}^{\lceil\log(1/\delta)\rceil}2^l/\eps = (2^{\lceil\log(1/\delta)\rceil+1}-1)/\eps \in \mathcal{O}(1/\delta_w \eps)$
\end{proof}

\paragraph*{Algorithm to place horizontal and large jobs} 

To place the jobs in $\Rho\cup \Rla$, we first guess the starting positions of the large jobs $\Rla$ in $\mathcal{O}(|S_{hor,large}|^{|\Rla|}) = (1/\eps)^{(1/\eps)^{\Oh(1/\eps)}}$. 
Note that this guess affects the energy demand that is left for horizontal jobs.
Next we guess which  $\mathcal{O}(1/\eps\delta)$ starting points in $S_{hor,large}$ will be used after the shifting due to Claim \ref{clm:StartingPintsHorizontalImproved}. 
There are at most $|S_{hor,large}|^{\mathcal{O}(1/\eps\delta)} = (1/\eps)^{(1/\eps)^{\Oh(1/\eps)}}$ possible guesses total.
We call the set of guessed starting points $\bar{S}_{hor,large}$.
For each starting point in $\bar{S}_{hor,large}$, we calculate the residual total energy demand, that is left after the guess for the large jobs. 
For a given $s \in \bar{S}_{hor,large}$ let $\height_{s,hor}$ be this residual total energy demand.
Consider the following linear program for horizontal jobs:
\begin{align*}
\sum_{i \in {\jobs}_{hor}} \sum_{\substack{s' \in \bar{S}_{hor,large} \\ s' \leq s < s'+\width(i)}} x_{i,s}\height(i) &\leq \height_{s,hor} & \forall s \in \bar{S}_{hor,large}\\
\sum_{s \in S_{hor,large}} x_{i,s} &= \height(i) & \forall i \in \jobs_{hor}\\
x_{i,s} &\geq 0 & \forall s \in \bar{S}_{hor,large}, i \in \jobs_{hor}
\end{align*} 

A basic solution to this linear program has at most $|\bar{S}_{hor,large}| + |{\jobs}_{hor}| = \Oh(1/\eps\delta)$ non zero components. 
We can guess the non zero components in at most $(|\bar{S}_{hor,large}| \cdot |\bar{\jobs}_{hor}|)^{|\bar{S}_{hor,large}| + |\bar{\jobs}_{hor}|} = (1/\eps)^{(1/\eps)^{\Oh(1/\eps)}}$.
Furthermore, we can guess their value with precision $\mu T$ in at most $(1/\mu)^{|\bar{S}_{hor,large}| + |\bar{\jobs}_{hor}|} = (1/\eps)^{(1/\eps)^{\Oh(1/\eps)}} $ guesses. 
Scheduling all the horizontal jobs integral and the error due to the precision add at most $2\mu T \cdot (|\bar{S}_{hor,large}| + |\bar{\jobs}_{hor}|)$ to the peak energy demand.
Note that $2\mu T \cdot (|\bar{S}_{hor,large}| + |\bar{\jobs}_{hor}|) \leq \Oh(\eps) T'$ since $\mu \leq \Oh(\eps^2\delta)$.

After this step, we either have scheduled all given jobs or have decided that it is not possible for the given guess of $T$ and the profile. 
If it is not possible for any profile, we have to increase $T$. 
If we have found a schedule, we try the next smaller value for $T$.
Each of the steps has increased the peak energy demand by at most $\Oh(\eps)T$ above $T$.
Besides of the job classification and rounding, each step of the algorithm is bounced by $(1/\eps)^{1/\eps^{\Oh(1/\eps)}}$.
Therefore, the described algorithms fulfills the claims of Theorem \ref{thm:aeptas}.

\section{Conclusion}
In this paper, we presented an AEPTAS with additive term $\hmax$ as well as a $(5/3+\eps)$-approximation for \acf{nonpreemptiveProblem}. 
Since the lower bound for approximation algorithms for this problem is known to be $3/2$, this leaves a small gap between the lower bound and the approximation guarantee. 
Closing this gap is an interesting open question for further research, especially since for the related strip packing problem the same gap is yet to be resolved. 

\bibliographystyle{plain}
\bibliography{bib}

\end{document}